\documentclass[a4paper,fleqn]{cas-sc}

\usepackage[numbers,sort&compress]{natbib}
\usepackage{graphicx}
\usepackage{amsmath}
\usepackage{soul} 

\usepackage{siunitx}
\usepackage{acro}
\acsetup{first-style=short}

\DeclareAcronym{AI}{
  short = AI ,
  long  = Artificial intelligence ,
  class = abbrev
}

\DeclareAcronym{CNN}{
  short = CNN ,
  long  = Convolutional neural network ,
  class = abbrev
}

\DeclareAcronym{DNN}{
  short = DNN ,
  long  = Deep neural network ,
  class = abbrev
}

\DeclareAcronym{ECFP}{
  short = ECFP ,
  long  = Extended connectivity fingerprint ,
  class = abbrev
}

\DeclareAcronym{GPCR}{
  short = GPCR ,
  long  = G protein-coupled receptor ,
  class = abbrev
}

\DeclareAcronym{GPU}{
  short = GPU ,
  long  = Graphics processing unit ,
  class = abbrev
}

\DeclareAcronym{GRU}{
  short = GRU ,
  long  = Gated recurrent unit ,
  class = abbrev
}

\DeclareAcronym{GNN}{
  short = GNN ,
  long  = Graph neural network ,
  class = abbrev
}

\DeclareAcronym{QSAR}{
  short = QSAR ,
  long  = Quantitative structure-activity relationship ,
  class = abbrev
}

\DeclareAcronym{LSTM}{
  short = LSTM ,
  long  = Long short-term memory ,
  class = abbrev
}

\DeclareAcronym{MLP}{
  short = MLP ,
  long  = Multilayer perceptron ,
  class = abbrev
}

\DeclareAcronym{NN}{
  short = NN ,
  long  = Neural network ,
  class = abbrev
}

\DeclareAcronym{RNN}{
  short = RNN ,
  long  = Recurrent neural network ,
  class = abbrev
}

\DeclareAcronym{SVM}{
  short = SVM ,
  long  = Support vector machine ,
  class = abbrev
}

\DeclareAcronym{VS}{
  short = VS ,
  long  = Virtual screening ,
  class = abbrev
}

\DeclareAcronym{VAE}{
  short = VAE ,
  long  = Variational autoencoder ,
  class = abbrev
}

\def\tsc#1{\csdef{#1}{\textsc{\lowercase{#1}}\xspace}}
\tsc{WGM}
\tsc{QE}
\tsc{EP}
\tsc{PMS}
\tsc{BEC}
\tsc{DE}

\begin{document}
\let\WriteBookmarks\relax
\def\floatpagepagefraction{1}
\def\textpagefraction{.001}
\shorttitle{AI-based GPCR bioactive ligand discovery}

\shortauthors{Raschka and Kaufman}

\title [mode = title]{Machine learning and AI-based approaches for bioactive ligand discovery and GPCR-ligand recognition}

\author[1]{Sebastian Raschka}[orcid=0000-0001-6989-4493]
\cortext[cor1]{Corresponding author}

\ead{sraschka@wisc.edu}
\ead[url]{http://pages.stat.wisc.edu/~sraschka/}
\fntext[fn1]{Medical Science Center, 1300 University Ave, Madison, 53706 WI}

\author[2]{Benjamin Kaufman}
\address[1]{University of Wisconsin--Madison, Department of Statistics}
\address[2]{University of Wisconsin--Madison, Department of Biostatistics and Medical Informatics}

\begin{abstract}
In the last decade, machine learning and artificial intelligence applications have received a significant boost in performance and attention in both academic research and industry. The success behind most of the recent state-of-the-art methods can be attributed to the latest developments in deep learning. When applied to various scientific domains that are concerned with the processing of non-tabular data, for example, image or text, deep learning has been shown to outperform not only conventional machine learning but also highly specialized tools developed by domain experts. This review aims to summarize AI-based research for GPCR bioactive ligand discovery with a particular focus on the most recent achievements and research trends. To make this article accessible to a broad audience of computational scientists, we provide instructive explanations of the underlying methodology, including overviews of the most commonly used deep learning architectures and feature representations of molecular data. We highlight the latest AI-based research that has led to the successful discovery of GPCR bioactive ligands. However, an equal focus of this review is on the discussion of machine learning-based technology that has been applied to ligand discovery in general and has the potential to pave the way for successful GPCR bioactive ligand discovery in the future. This review concludes with a brief outlook highlighting the recent research trends in deep learning, such as active learning and semi-supervised learning, which have great potential for advancing bioactive ligand discovery. 
\end{abstract}

\begin{keywords}
molecular representations \sep GPCR ligands \sep drug discovery \sep
deep learning \sep machine learning \sep graph convolutional neural networks \end{keywords}

\maketitle

\newpage

\printacronyms[include-classes=abbrev,name=Abbreviations]

\section{Introduction}
\label{sec:introduction}

G protein-coupled receptors (\ac{GPCR}s) are one of the most prominent families of integral membrane proteins and are among the most widely studied targets in drug discovery and development. According to~\cite{hauser2017trends}, 34 percent of all drugs approved by the US Food and Drug Agency target GPCRs. While the exact size of the GPCR family remains to be determined, genome analyses suggest that the GPCR family is represented by approximately 800-1,000 genes in humans~\cite{garland2013gpcrs,thomsen2005functional}. More than 150 GPCRs are characterized as orphan receptors, which means that the receptors' endogenous ligands are still unknown~\cite{bjarnadottir2006comprehensive}. However, it remains unclear whether endogenous ligands exist for all orphan GPCRs~\cite{davenport2013international}.   

Since GPCRs are involved in many different cellular and biological processes, make excellent drug targets, and remain orphaned to a relatively large extent, the prediction and consequent identification of GPCR ligands is an active area of research and interest~\cite{raschka2019automated}. Recent studies suggest that drugs target only approximately 10\% of known GPCRs~\cite{southan2015iuphar}. The high cost of clinical trials coupled with a relatively low success rate (approximated to be below 6.2\%~\cite{wong2019estimation,vamathevan2019applications}), motivates the development and application of machine learning and artificial intelligence-based methods for ligand discovery, to lower costs as well as to identify candidates that would otherwise be missed using conventional methods.  

\subsection{GPCRs and drug discovery}
\label{sec:gpcrs-and-drug-discovery}

Many common human diseases involve GPCR signaling, including schizophrenia, glaucoma, depression, and hypertension~\cite{garland2013gpcrs}. Despite the increasing popularity of biologics, which includes gene therapy, recombinant therapeutic proteins, antibody-drug conjugates, and tissues~\cite{biologics2019web}, experts see the discovery of small-molecule ligands being crucial for the future of molecular medicine and the treatment of human diseases~\cite{mullard2019fda,rodrigues2020machine}.  

GPCR ligands exhibit a wide range of characteristics and are very diverse in their physiochemical properties, shape, and size~\cite{southan2015iuphar}. Furthermore, GPCR ligands include lipids, peptides, proteins, steroids, and other small organic molecules. Depending on their mechanism of action, GPCR ligands can be described as full or partial agonists, antagonists, or inverse agonists. This diversity among GPCR ligands poses challenges for standardized binding assays used in experimental bioactivity screening. Thus, wet-lab techniques such as high throughput screening have risen in popularity. However, high throughput screening can still be cost-prohibitive and labor-intensive. Another downside of high throughput screening is the limitation of available ligand libraries in terms of size and diversity. With drug discovery being an expensive and labor-intensive process, with estimated costs ranging between 0.5-2.6 billion US dollars, and a timeline ranging between 10-20 years, researchers have begun to favor computational methods for identifying candidate molecules for more elaborate bioactivity assays~\cite{paul2010improve,zhavoronkov2019deep}.  

When it comes to the identification of protein targets of small-molecule ligands and assessing potential off-target activity, chemical proteomics are considered the gold standard~\cite{parker2017ligand,rodrigues2020machine,bar2017chemical,moellering2012chemoproteomics}. However, similar to the challenges of binding assays for small molecule screening, chemical proteomics experiments are laborious and challenging to streamline~\cite{laraia2017natural}.  Hence, many researchers are increasingly favoring computational alternatives~\cite{rodrigues2020machine} or augmenting the experimental discovery pipeline with artificial intelligence~\cite{duros2017human,hase2019next}.   

\subsection{Computer-aided ligand discovery}
\label{sec:computer-aided}

\begin{figure}
\centering
\includegraphics[width=0.8\textwidth]{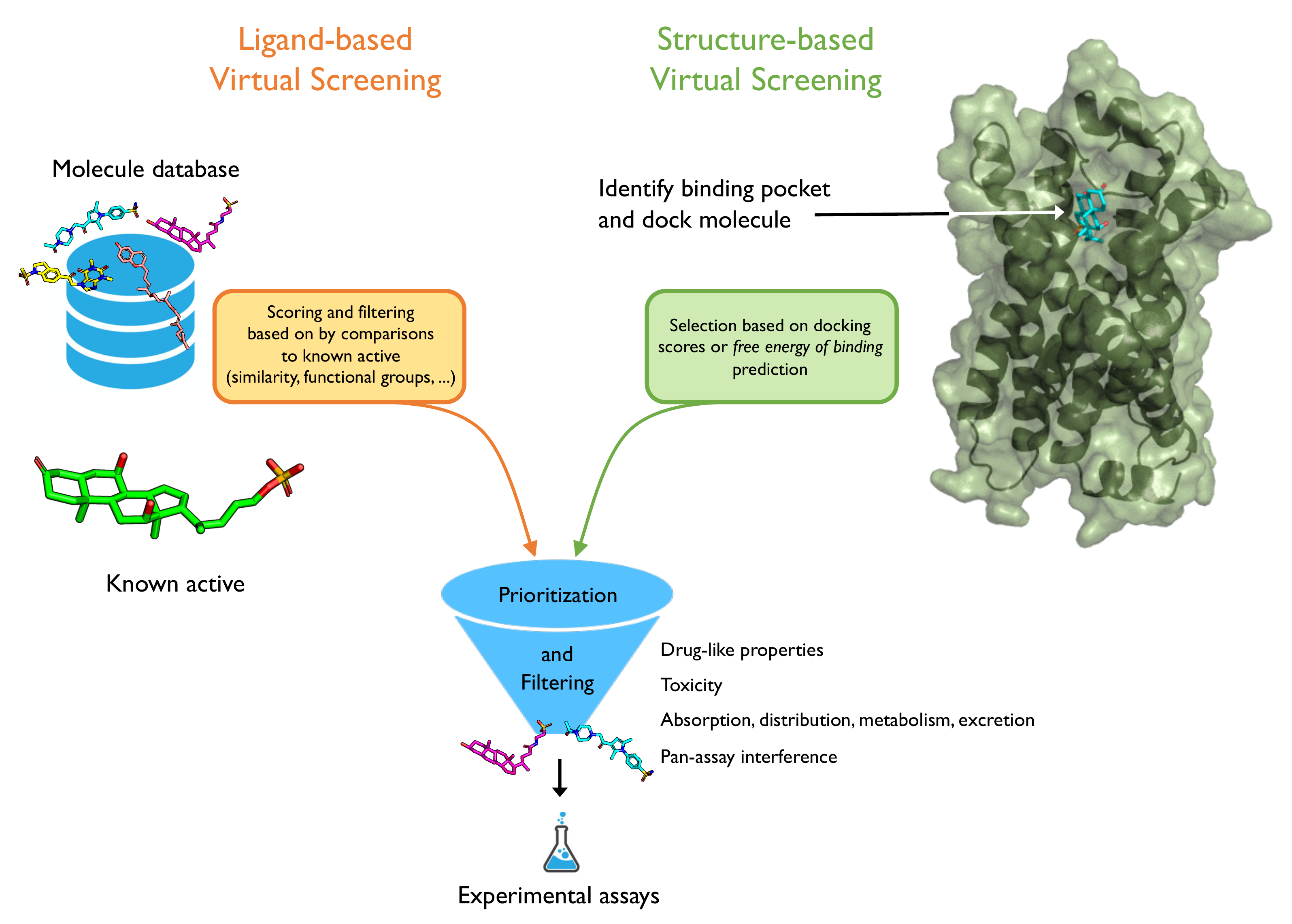}
\caption{{Conceptual overview of ligand-based and structure-based virtual screening depicting 3kPZS (known active), which is a pheromone compound involved GPCR signaling, and a homology structure of its receptor, the SLOR1 GPCR~\cite{raschka2018enabling}.}
}
\label{fig:vs}
\end{figure}

Modern computational databases feature hundreds of millions of molecules and are available free of charge~\cite{sterling2015zinc,sunseri2016pharmit,bento2014chembl}, which makes computer-aided ligand discovery a particularly compelling alternative to high throughput screening and other experimental approaches during the early stages of ligand discovery. The two major approaches for computer-aided ligand discovery, also known as virtual screening (VS), are ligand-based VS and structure-based VS {Figure~\ref{fig:vs}.}

{\subsubsection{Ligand-based VS}}
\label{sec:ligand-based-vs}
Ligand-based VS approaches focus on the structure and physicochemical properties of ligands in the absence of the receptor structure. Broadly, ligand-based VS can be described as a set of techniques for ligand-based similarity search or property prediction. Ligand-based VS approaches for identifying bioactive molecules are often based on a known bioactive molecule given the hypothesis that similar molecules are likely to bind the target receptor and exhibit a certain level of bioactivity, which has to be determined experimentally. {As traditional ligand-based VS depends on pair-wise similarity measures between database molecules and the known active, researchers have to choose a meaningful similarity or distance measure. The choice of an appropriate similarity measure, in turn, depends on the molecular representation, which we discuss in more detail in Section~\ref{sec:molecular-representation}.}

{To provide an illustrative example of the use of ligand-based VS in GPCR bioactive ligand discovery, we consider the recent discovery of a potent inhibitor of a GPCR-mediated pheromone signaling pathway in sea lamprey~\cite{raschka2018enabling}. Based on volumetric and physicochemical overlays between 3D conformers of database molecules and the cognate ligand, researchers were able to identify a small molecule capable of blocking a GPCR-mediated signaling response. However, when the researchers assayed the activity of the 299 top-scoring molecules, they found no correlation between molecular activity and the 3D overlay-based similarity scores. Additionally, the researchers found that changing only a single chemical group (for instance, keto to hydroxyl), which may only have a small impact on the overall similarity score, can completely suppress GPCR-mediated signaling in the pheromone pathway. This finding provides supportive evidence that general similarity measures are useful but not sufficient means for active molecule discovery. However, the virtual screening data combined with the activity measurements collected via experimental assay data offer excellent opportunities to utilize machine learning for developing custom activity classifiers and conducting quantitative structure-activity relationship (\ac{QSAR}) analyses. For instance, in a follow-up study, the researchers described how machine learning techniques were used to identify the chemical features that correlated with bioactivity~\cite{raschka2018automated}. To facilitate this analysis, the researchers converted functional group matches between database compounds and the known active into binary feature vectors. For example, if the keto-group of the known active was within 1.3 \si{\angstrom} of a query molecule's keto-group (in the in the 3D overlay), it was counted as a functional group match (1) and as a non-match (0), otherwise. After tabulating the functional group matches as binary feature vectors, the researchers fit logistic regression and random forest classification models and quantified feature importance of the functional and chemical groups. The importance of certain chemical groups was then used to inform subsequent rounds of virtual screening, to identify database molecules that were previously dismissed in early stages of the ligand-based VS pipepline.
}

{\subsubsection{Receptor structure-based VS}}
In contrast to ligand-based VS, receptor structure-based VS assumes knowledge of the receptor structure with molecular docking being one of the most prominent structure-based VS techniques. While molecular docking has led to many successful discoveries, one of its limitations for GPCR ligand discovery is the small number of high-resolution GPCR structures that are currently available. Recent years brought many significant improvements for protein X-ray crystallography and cryo-electron microscopy; however, high-resolution structures still only cover four out of the six different classes of GPCR: classes A, B, C, and F, of which the rhodopsin-like receptors from class A form the most significant portion~\cite{basith2018exploring}. Furthermore,  another limitation for structure-based approaches for bioactive ligand discovery is the breadth and diversity of GPCR structures that are currently available: only 44 of the 205 GPCR structures correspond to unique GPCRs~\cite{hauser2017trends}.  

All GPCRs consist of seven transmembrane helices, which are relatively conserved across the different classes of GPCRs (A-F). However, GPCRs may differ more noticeably in the intracellular and extracellular loop domains. The extracellular loops play an essential role in many GPCR structures, since they often form the orthosteric ligand binding site or provide access to binding sites that are located within the transmembrane bundle~\cite{wheatley2012lifting}. The diversity of GPCR ligands and GPCR ligand binding sites are a direct consequence of the structural variety of the extracellular loops, which makes employing structure-based approaches extremely challenging in the absence of high-quality structural information. In those cases, ligand-based strategies may represent the only viable alternative~\cite{raschka2018enabling}. In recent years, both traditional structure-based and ligand-based VS have been augmented or replaced by machine learning techniques. Furthermore, machine learning has also been used for predicting bioactivity from other types of data such as a bioactivity matrix (a targets-by-compounds matrix of functional interactions)~\cite{zhang2019biomatrix}. Today, machine learning is widely recognized as an essential method in the chemical biology toolbox for researching ligand binding~\cite{rodrigues2020machine}. 

{Morevover, besides binding mode prediction and scoring, there are various other aspects of structure-based approaches that can benefit from machine learning. For instance, in a recent study, researchers combined X-ray crystal structural analysis with machine learning to identify key features distinguishing active from inactive states of class A GPCRs that were induced by bioactive ligands~\cite{bemister2020machine}. The researchers started with rigidity analysis~\cite{jacobs2001protein} to obtain rigidity information for a set of ligand-bound and ligand-free GPCR crystal structures in active and inactive states. Next, the researchers partitioned the helical regions of the receptors into 29 segments, where each segment was subdivided into 3 blocks. The 87 blocks where then used to construct feature vectors as input to a machine learning classifier. In particular, each GPCR was represented by a continuous feature vector containing 87 values, each value being a rigidity index (measured on a scale ranging from 0 to 100) for that block. After assembling this tabular dataset, the researchers trained a k-nearest neighbor classifier that was able to predict active and inactive states with high (96\%) accuracy. In addition, the researchers gained further insights, namely the identification of six key flexibility transition regions, which are involved in activating (or inactivating) the GPCR upon ligand binding. We hypothesize that if protein-ligand interactions are known or can be reliably predicted then this type of analysis can aid virtual screening of bioactive GPCR ligand as well as GPCR ligand design.}

\vspace{1em}
\subsection{Augmenting ligand discovery with artificial intelligence and machine learning}
\label{sec:augmenting}

Artificial intelligence (\ac{AI}) is traditionally considered a subfield of computer science that focuses on tasks that humans are naturally good at, such as image recognition and natural language processing. Furthermore, {AI} can be categorized into artificial general intelligence, i.e., human-level intelligence on a breadth of different tasks, and "narrow" AI, which focuses on solving a single task well. Examples of narrow AI include tasks such as classifying objects from images, language translation, playing chess, or predicting the bioactivity of small molecules. Currently, the most popular approach for implementing and programming narrow AI is machine learning. Machine learning is a field that focuses on the development of algorithms that allow computers to learn from representative datasets, as opposed to having domain experts developing rules manually.  

The three major subcategories of machine learning are supervised learning, unsupervised learning, and reinforcement learning. In supervised learning, the goal is to predict a category label (classification) or score (regression analysis) by learning from a large collection of such labeled examples. In other words, supervised learning is concerned with learning a mapping function between the so-called input features or observations (for example, fingerprint representations of small molecules) and a discrete or continuous target variable, for example, an active/inactive label or the binding affinity in the context of a specific receptor. After learning to predict the desired scores or labels from the labeled training dataset, an independent test dataset, consisting of labeled examples that were unseen during training, is used to evaluate the performance of the predictive model. 

\begin{figure}
\centering
\includegraphics[width=0.95\textwidth]{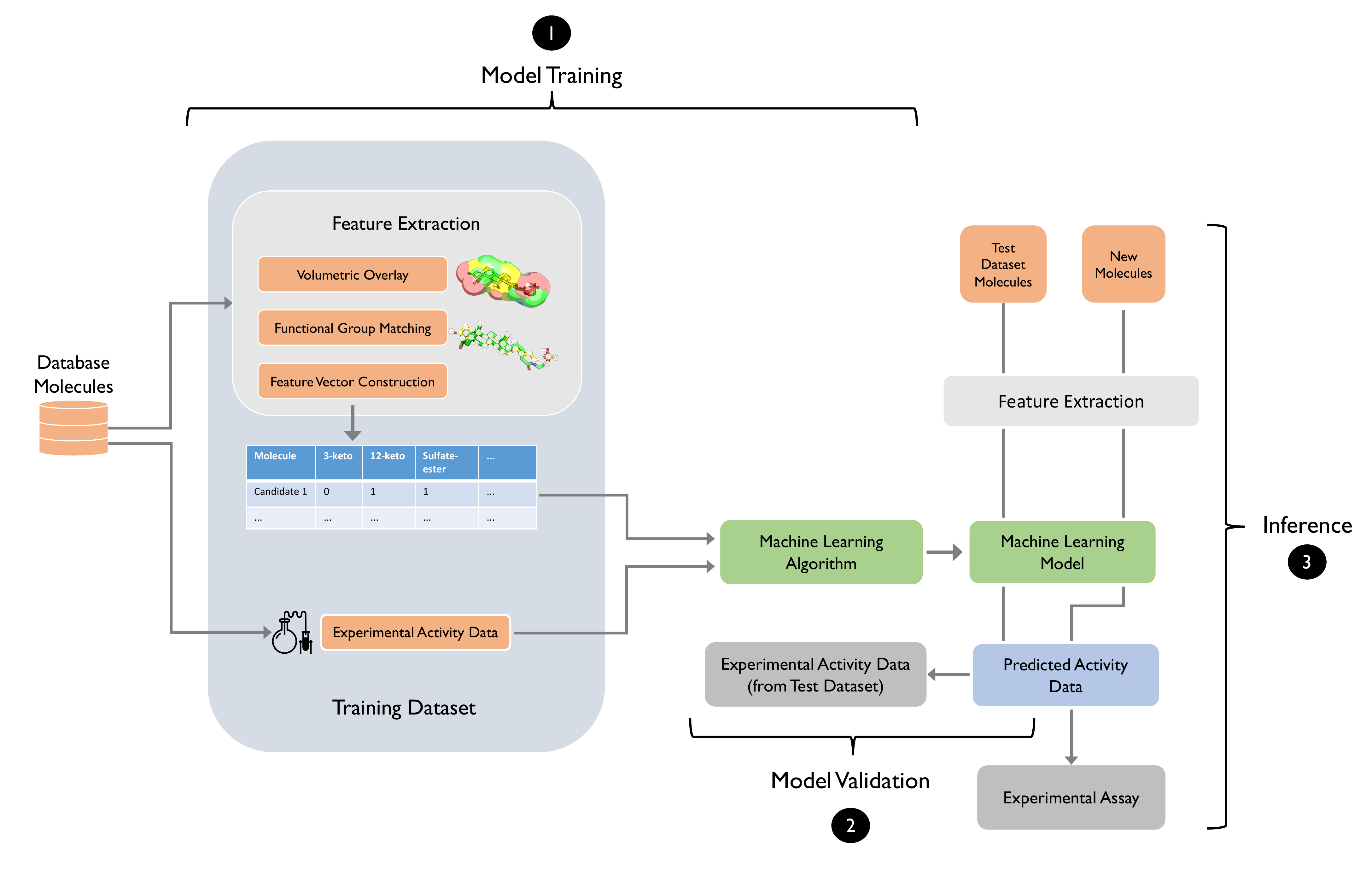}
\caption{{Illustration of a supervised learning workflow for bioactive ligand discovery summarizing the GPCR inhibitor discovery described in ~\cite{raschka2018enabling} and~\cite{raschka2018automated}. Step 1: After extracting feature vectors for each candidate molecule, the training dataset consisting of bioactivity measurements and the feature vectors is used to train the machine learning model. Step 2: The model is evaluated on the test dataset, which contains molecules (and their experimental bioactivity measurements) to validate the model. Step 3: If the model yielded satisfactory validation performance, it could be used to predict the activity of new molecules and prioritize candidates for experimental assays.}}
\label{fig:supervised-workflow}
\end{figure}

{To illustrate these supervised learning concepts in the context of GPCR bioactive ligand discovery, we revisit the SLOR1 receptor signaling inhibitor projects~\cite{raschka2018enabling,raschka2018automated} introduced in Section~\ref{sec:ligand-based-vs}. Suppose the goal is to train a classification model to predict whether a (new) candidate molecule is active against SLOR1. To train such a classifier, we require a labeled dataset, which contains the target variable we want to predict. The target variable, in this case, is a binary variable with two possible values, \textit{active} and \textit{non-active}. Ideally, the target labels are obtained from experimental assays, for instance, by assaying a set of candidate molecules from a molecular database for bioactivity against the target receptor. (Note that it is also possible to model this prediction problem as a regression task, where the target variable is a continuous activity value; however, in our experience, classification models are less sensitive to noise, easier to fit, and more robust overall.) Next, we have to decide about the feature representation of the molecules that are presented to the machine classifier along with the target labels. For example, we may use the functional group matching vectors based on 3D volumetric overlays with a known active as described in Section~\ref{sec:ligand-based-vs} and~\cite{raschka2018automated}. Before we start training the classifier, we divide the dataset into a separate training and test set, where the training set is used to fit the model. After model fitting, the test set can be used to evaluate the classifier's performance using molecules that it has not seen during training. During the evaluation, the model is only presented with the feature vector representation of the molecules. The predicted activity labels are then compared to the experimentally measured activity levels to quantify predictive performance. Since model evaluation is an important aspect of machine learning, which we cannot cover in detail here, we recommend consulting the article "Model evaluation, model selection, and algorithm selection in machine learning"~\cite{raschka2018model} to learn more about the best practices. If the model has achieved satisfactory performance, we may use it on a new database of molecules to predict whether they are likely to be active or not. This process is summarized in Figure~\ref{fig:supervised-workflow}.}

Unsupervised learning is used for tasks that do not involve target labels. A typical example of unsupervised learning is clustering, for instance, grouping small molecules based on a user-specified similarity measure. 

Reinforcement learning, a third major category of machine learning, focuses on decision making in complex environments, which requires learning an optimal series of actions in response to environmental information as opposed to a target output as in supervised learning. A classic example is the development of a chess program.  

In the presence of labeled high-quality data, machine learning provides opportunities for automating the development of customized solutions that are optimized for a specific type of receptor, binding site, or small molecule -- as opposed to more general scoring functions such as DrugScore~\cite{neudert2011dsx} or AutoDock Vina~\cite{trott2010autodock}. Machine learning has already  shown to be successful in various stages throughout typical ligand and drug discovery, including the discovery of novel drug targets and ligands~\cite{jeon2014systematic,riniker2014using}, bioactivity prediction~\cite{rifaioglu2018recent}, predicting new binding sites in GPCRs~\cite{chan2019new,wu2018coach}, analyzing the association between the drug targets and diseases~\cite{ferrero2017silico}, studying binding pathways (for example, the binding pathways of opiates to $mu$-opioid receptors~\cite{farimani2018binding}), optimizing lead compounds~\cite{ballester2010machine}, molecular {\it de novo} design~\cite{olivecrona2017molecular}, modifying molecular properties~\cite{kadurin2017drugan}, and developing biomarkers to assess the efficacy of drugs~\cite{mamoshina2018machine}.

\subsection{Deep learning--representation learning with deep neural networks}
\label{sec:representation-learning}

In the last decade, deep learning research has seen a substantial increase in attention, since it has allowed researchers to develop state-of-the-art solutions for computer vision, natural language processing, and computational biology~\cite{lecun2015deep}. Deep learning is a subfield of machine learning that focuses on artificial neural networks with many layers that can learn multiple levels of abstractions of data that are useful for supervised and unsupervised learning tasks. Modern deep learning architectures consist of hundreds of millions of parameters~\cite{canziani2017analysis}, the so-called model weights, which are learned from a training dataset using the backpropagation algorithm~\cite{rumelhart1986learning}. An overview of the most commonly used deep learning architectures is provided in Section~\ref{sec:dl-architectures}.   

The predictive performance of traditional machine learning methods -- this includes logistic regression, random forests, support vector machines (\ac{SVM}s), k-nearest neighbors, and many others -- heavily depends on the design of the feature engineering pipeline. For instance, the aforementioned conventional machine learning methods generally do not operate well on high dimensional datasets and are unable to extract knowledge from raw data (such as text or images)~\cite{lecun2015deep}.  {Hence, the training data has to be converted into a tabular format via manual feature engineering, where \textit{manual} means that the feature extraction is not implicitly realized by the model but has to be carried out beforehand via an additional procedure. For example, in a virtual screening context, the two- or three-dimensional graph representations of a molecule represent \textit{raw} data, whereas simplified molecular-input line-entry system (SMILES) string or fingerprint representations of a molecule are the outcomes of manual feature extraction. Different types of molecular representations will be covered in more detail in Section~\ref{sec:molecular-representation}.}

Careful feature engineering requires substantial domain expertise, and useful information contained in the data can be lost by feature engineering. Deep learning, on the other hand, relies on general-purpose algorithms that include the automatic extraction of salient information from raw data as part of the modeling architecture and optimization objective. In this respect, deep learning can be characterized as a feature or representation learning method. However, one downside of deep learning is that it is relatively data-hungry, and datasets for supervised learning require large amounts of labeled examples -- dataset sizes ranging between 50 thousand and 15 million training examples are not unusual~\cite{cao2019consistent,deng2009imagenet}. 

\subsection{Open source software and datasets} 
\label{sec:open-source-software} 

Over the past decade, there has been a remarkable increase in the development of open source software for enabling data science and machine learning~\cite{raschka2019python}. Many general-purpose scientific computing libraries with permissive open source licenses are now widely used in academia as well as in industry, including NumPy and SciPy~\cite{van2011numpy}, Matplotlib~\cite{hunter2007matplotlib}, and Pandas~\cite{mckinney2010data}. Furthermore, general-purpose libraries have been developed for the analysis of biological and structural data that lower the barrier of entry to computational biology, for example, BioPython~\cite{cock2009biopython} and BioPandas~\cite{raschka2017biopandas}.   

Under similar open source licensing terms, the Scikit-learn machine learning library~\cite{pedregosa2011scikit} has been widely adopted for predictive modeling with traditional machine learning (for example, generalized linear models, SVMs, and tree-based methods such as random forests and gradient boosting). In recent years, computationally efficient and \ac{GPU}-enabled libraries such as TensorFlow~\cite{abadi2016tensorflow} and PyTorch~\cite{paszke2019pytorch} made deep learning accessible to the broad scientific communities similar to how Scikit-learn contributed to popularizing machine learning.  

While the use of machine learning and deep learning has seen widespread adoption throughout various research areas, a bottleneck for applications to computational biology was the lack of large, annotated, high-quality datasets. However, recent years brought both improvements of experimental techniques and the "data and code sharing" culture in academia, which led to the increase of publicly available molecular activity and biomedical datasets. For instance, in drug discovery, recent efforts focused on developing an open, annotated dataset that can be utilized for therapeutic target validation~\cite{koscielny2016open}. MoleculeNet is a large benchmark dataset consisting of more than 700,000 molecules and their property annotations~\cite{wu2018moleculenet}. The ChEMBL large-scale bioactivity database contains more than 5.4 million bioactivity measurements for 5,200 protein targets and more than 1 million ligands~\cite{gaulton2016chembl}. While the aforementioned databases are general-purpose datasets that are usually used for general algorithm development and benchmarking, they can also be used as a starting point for pre-training deep learning models that can then be fine-tuned to smaller datasets for specific targets. This technique is called transfer learning and is discussed in greater detail in Section~\ref{sec:transfer-learning}. 

\subsection{Interpretability and repeatability of machine learning}

Machine learning models can learn to identify salient patterns in high-dimensional and complex datasets that are not obvious to a human researcher. However, machine learning and particularly deep learning models are often criticized for their black box aspect and lack of interpretability~\cite{vamathevan2019applications}. We agree that machine-learning-based methods, except for the most fundamental decision tree models and rule-based classifiers, are less interpretable than pure "if/then" rules. However, many methods exist that allow scientists to obtain insights into which features a model learns for making particular predictions. {For example, in the SLOR1 context discussed in Sections~\ref{sec:ligand-based-vs} and~\ref{sec:augmenting},} the researchers were able to identify a potent inhibitor of GPCR pheromone receptor signaling using a machine learning-aided virtual screening approach that provided structure-activity information~\cite{raschka2018enabling}. Furthermore, by combining machine learning with feature selection algorithms, the researchers were able to identify sulfate groups on the tail end of the candidate ligands that are crucial for bioactivity in SLOR-1 GPCR signaling~\cite{raschka2018automated}. While the methods described in~\cite{raschka2018automated} were applied to summarize structure-activity relationships of 3D-structural overlays of small molecules that were prioritized for experimental bioassays, the same techniques can be used for different types of data, such as hydrogen bonding patterns~\cite{raschka2018protein} or ligand-receptor docking poses~\cite{raschka2016detecting}.

\begin{figure}
\centering
\includegraphics[width=0.9\textwidth]{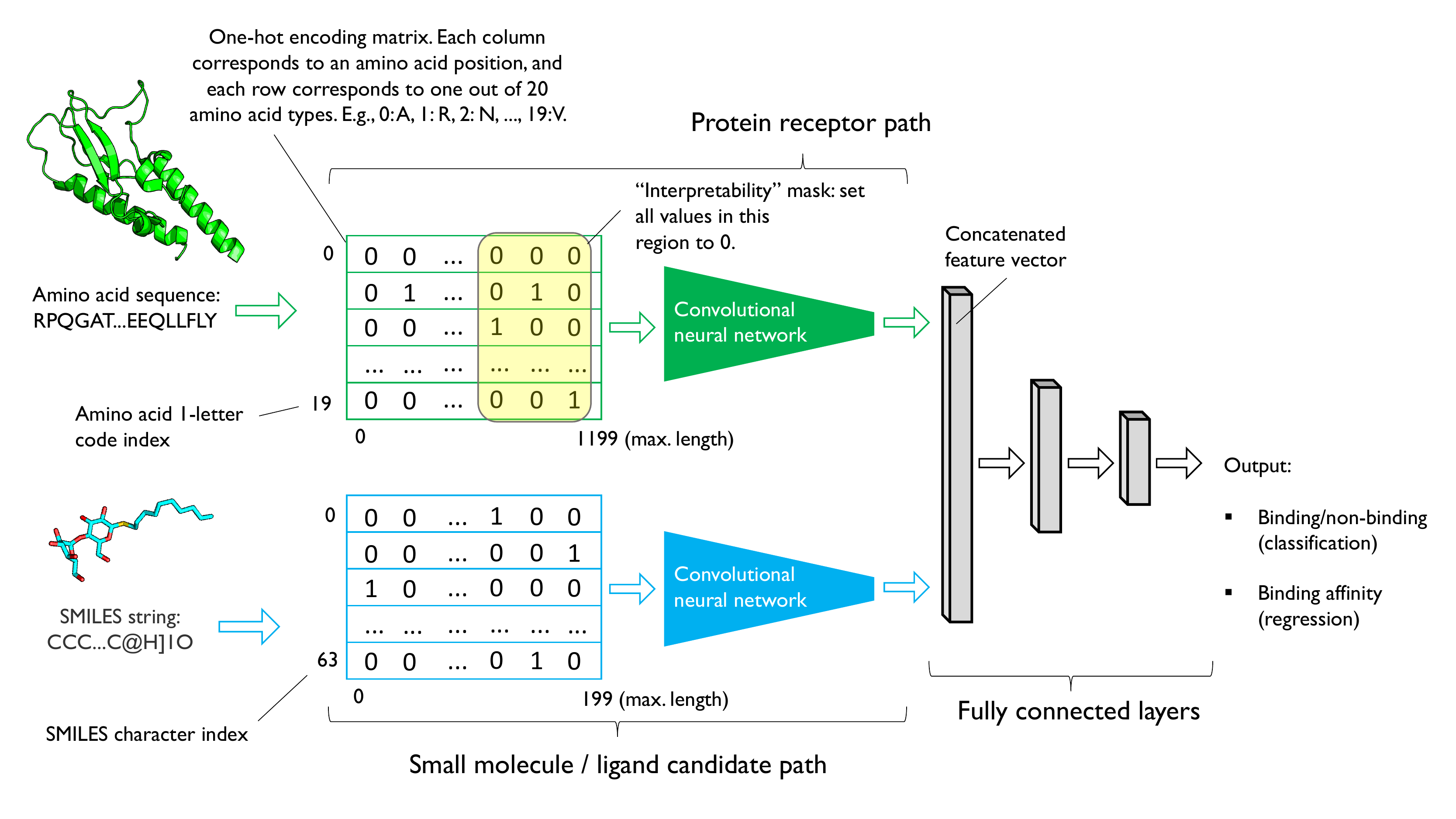}
\caption{{Illustration of a convolutional neural network trained on 1D representations of protein receptors and ligand candidates. The column vectors in the one-hot encoded matrix are sparse, consisting of one "1" (indicating the type of amino acid or SMILES character at a given position), and the remaining values are "0"'s. After training, an "interpretability" mask, which sets all values in the masked region to 0, can be used to observe how blocking out subsequences affects the binding prediction. (The crystal structure corresponds to the ligand-bound glucagon-like peptide-1 receptor extracellular domain~\cite{runge2008crystal}, PDB code: 3C5T.)}
}
\label{fig:interpretable-cnn-1}
\end{figure}

Other commonly used methods for understanding model predictions are general-purpose methods such as LIME~\cite{ribeiro2016should} or SHAP~\cite{lundberg2017unified}. Neural network-specific methods include guided backpropagation~\cite{springenberg2014striving}, class activation mapping~\cite{zhou2016learning}, gradient-weighted class activation mapping~\cite{selvaraju2017grad}, and learning important features through propagating activation differences (DeepLIFT)~\cite{shrikumar2017learning}. {All of these methods can be used to study how changes in the inputs affect the model's output (predictions). Hu et al.~\cite{hu2019interpretable} demonstrated how a simple variant of this concept could be applied to identify binding sites from 1D sequence data. Assuming that 3D structures are unavailable or too costly to obtain, the researchers trained deep neural networks on 1D representations of proteins (1-letter amino acid codes) and ligands (SMILES strings) to predict binding information. In particular, the researchers trained convolutional neural networks (see Section~\ref{sec:cnns}) for binary classification (binding vs. non-binding) and regression (binding affinity) on one-hot encoded versions of the 1D sequences as illustrated in Figure~\ref{fig:interpretable-cnn-1}. After model training and evaluation, parts of a protein sequence were masked to analyze how their removal impacted the prediction score. If the masking of a subsequence had a significant effect on the predicted output, one could hypothesize that the subsequence was important for binding.}

Another point of criticism against the use of deep neural networks is the lack of repeatability~\cite{vamathevan2019applications}. In this context, repeatability refers to the ability to produce the exact same results upon repeating an experiment under identical conditions; in contrast, reproducibility refers to the ability to obtain similar results in different environments or conditions, for example, if a different research lab conducts similar experiments. According to~\cite{vamathevan2019applications}, the issue of repeatability "arises because [machine learning] outputs are highly dependent on the initial values or weights of the network parameters or even the order in which training examples are presented to the network, as all of them are typically chosen at random." We want to highlight that these repeatability issues can easily be circumvented by specifying the seeds of pseudo-random number generators and setting the behavior of machine learning and deep learning software libraries to deterministic. In practice, however, issues with reproducibility and repeatability arise when researchers share insufficient details about the software versions that were used to conduct the experiments. Reproducibility and repeatability require best code sharing practices, which include not only sharing the exact code but also the experimental settings and software version numbers. Fortunately, sharing data and code has become a recommended practice in the field of machine learning~\cite{neurips2018callforpapers}, and we hope that computational biology journals will adopt similar best practices such that repeatability and reproducibility issues that plagued the field in previous years can be avoided in the future. 

\subsection{Commonly used deep learning architectures}
\label{sec:dl-architectures} 

This section summarizes the main concepts behind machine learning with a focus on the fundamental deep learning concepts that are relevant for the remainder of this article. For an introduction and overview of general machine learning methods, we recommend~\cite{burkov2019hundred,raschka2019python}. For a more comprehensive introduction to deep learning, we recommend~\cite{lecun2015deep,goodfellow2016deep,raschka2019python}.

\subsubsection{Multilayer perceptrons}

\begin{figure}
\centering
\includegraphics[width=0.95\textwidth]{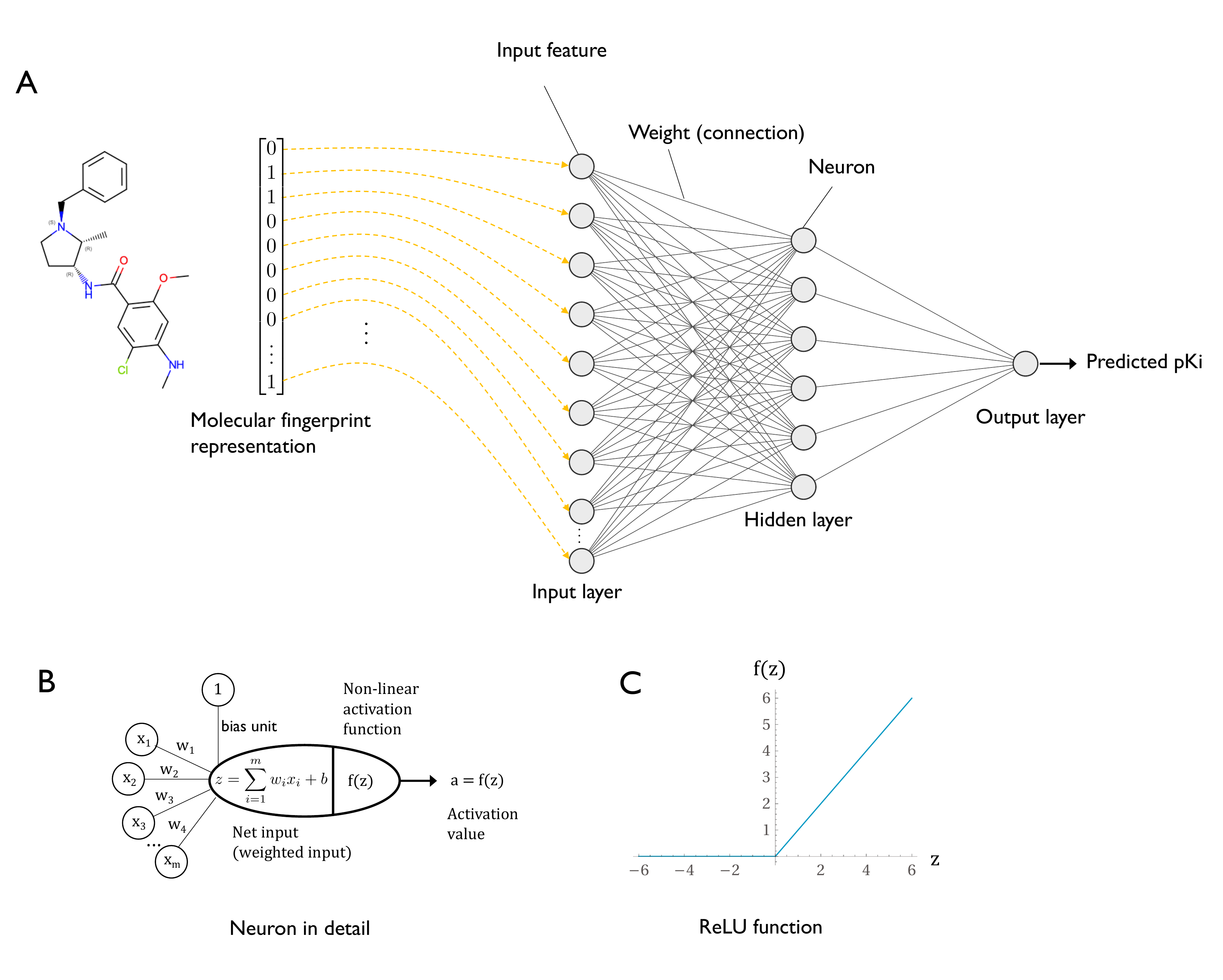}
\caption{{(A) A fully connected neural network (multilayer perceptron; MLP) with one hidden layer. The input layer represents the input features from the dataset, for instance, a molecular fingerprint representing a small molecule (here: 
Dopamine D4 receptor antagonist Nemonapride). The input layer is "fully connected" to the hidden layer via weight connections. In this example, the output layer consists of a single neuron that outputs a continuous value, the predicted $pK_i$. (B) Depiction of a hidden layer neuron in greater detail. The neuron computes the weighted sum of all input features (plus a bias or threshold value) to calculate the so-called net input. This weighted input is then passed to a non-linear activation function, such as the rectified linear unit (ReLU) depicted in (B).}}
\label{fig:mlp-1}
\end{figure}

Multilayer perceptrons (\ac{MLP}s) are fully connected artificial neural networks (\ac{NN}s) that consist of an input layer, an output layer, and at least one hidden layer between the two (Figure~\ref{fig:mlp-1} A). While there is no precise definition of what constitutes a \textit{deep} neural network (\ac{DNN}), an NN that has more than one hidden layer is commonly referred to as a DNN. The hidden units in the hidden layer manipulate the input information (observations or features) in a non-linear fashion so that, in the case of supervised classification, the training examples from different classes become linearly separable by the last layer. In DNNs, the early layers can be considered representation learning layers that distort the data in such a way that it can be classified by an output layer that has a similar structure to a generalized linear model. In other words, NN architectures with one or more hidden layers can learn highly complex relationships between the input features and the target label.   

An MLP with only one (sufficiently large) hidden layer can already be considered a universal function approximator~\cite{csaji2001approximation,cybenko1989approximations,hornik1989multilayer}. However, a DNN with many hidden layers can achieve the same expressiveness with fewer parameters than an NN with only one hidden layer. Furthermore, constructing an architecture with multiple layers provides some form of regularization: later layers are constrained by the behavior of earlier layers. Unfortunately, a common problem with deep architectures is that increasing the number of layers exacerbates the vanishing and exploding gradient problems that arise from repeated multiplications via the chain rule in backpropagation, making it hard to parameterize very deep models. A particular focus in the deep learning field has been on developing methods that help with training very deep architectures successfully, as discussed in Section~\ref{sec:cnns} and Section~\ref{sec:rnns}.

While the early MLP architecture was first proposed in the early 1960s~\cite{schmidhuber2015deep,ivakhnenko1966cybernetic,ivakhnenko1967cybernetics}, efficient ways for training such multilayer neural networks using the backpropagation procedure were formulated by several researchers independently between 1970 and 1986~\cite{linnainmaa1970representation,werbos1974beyond,rumelhart1986learning}. Although the backpropagation procedure is still the main learning algorithm in deep learning, the training of DNNs has only become broadly feasible in recent years due to several advances towards making the general training more efficient and effective. These improvements include implementing machine learning models and training DNNs on GPUs~\cite{steinkraus2005using,chellapilla2006high,raina2009large,cirecsan2012multi}, which are extremely well-suited for performing linear algebra operations such as matrix multiplications efficiently and on a large scale.  

In addition to developing software libraries that allow researchers to develop and implement DNNs more easily (see Section~\ref{sec:open-source-software}), countless algorithmic advances towards training DNNs more efficiently and robustly have been made in recent years. For example, new activation functions have been proposed, such as the rectified linear unit (ReLU; Figure~\ref{fig:mlp-1} C)~\cite{nair2010rectified}, which has become a popular default choice since it can help with vanishing gradient effects and generally allows for training DNNs faster and with better predictive performance. New stochastic gradient descent-based optimization algorithms such as Adam~\cite{kingma2014adam} can accelerate the model training and make it easier to find a proper learning rate setting compared to conventional stochastic gradient descent. Regularization methods such as Dropout~\cite{srivastava2014dropout} help reduce the effect of overfitting by dropping nodes randomly during training and thus make the model less reliant on particular inputs and connectivity patterns. Batch normalization is a method for scaling layer inputs~\cite{ioffe2015batch}, which can speed up learning further by allowing learning with larger batch sizes and converging to local minima on the loss surface in fewer iterations over the training set. While not all of these improvements exist in all modern architectures, they have been fundamental to allowing experts as well as non-experts to train DNNs successfully on a wide range of different datasets.  

{Koutsoukas et al.~\cite{koutsoukas2017deep} applied a multilayer perceptron model for diverse bioactivity prediction ($pK_i$ and $p\text{IC}_{50}$) against 7 different targets, including two GPCRs, dopamine receptor D4 and cannabinoid receptor 1 based on molecular fingerprint representation of the molecules (similar to the illustration in Figure~\ref{fig:mlp-1} A). The researchers also found that the MLP outperformed traditional machine learning approaches on large datasets. However, they noted that deep learning models require much more extensive hyperparameter tuning to achieve good predictive performance~\cite{koutsoukas2017deep}. An approach like this can be used for ligand-based discovery, when information about the binding interface is not available.}

{MLPs can also be utilized for structure-based VS or QSAR studies as described~\cite{ma2015deep}. Similar to~\cite{koutsoukas2017deep}, the researchers found that MLPs generally outperformed traditional machine learning methods like random forests and SVMs.  This is a remarkable result because, in this case, the model the researchers referred to as a DNN was a simple MLP with only one hidden layer. The descriptors used in this study are a union of atom pair and donor-acceptor pair descriptors, including information about atom types and binding interface information. Additional details about the different molecular data input representations will be provided in Section~\ref{sec:molecular-representation}.}

\subsubsection{Convolutional neural networks }
\label{sec:cnns}

\begin{figure}
\centering
\includegraphics[width=0.8\textwidth]{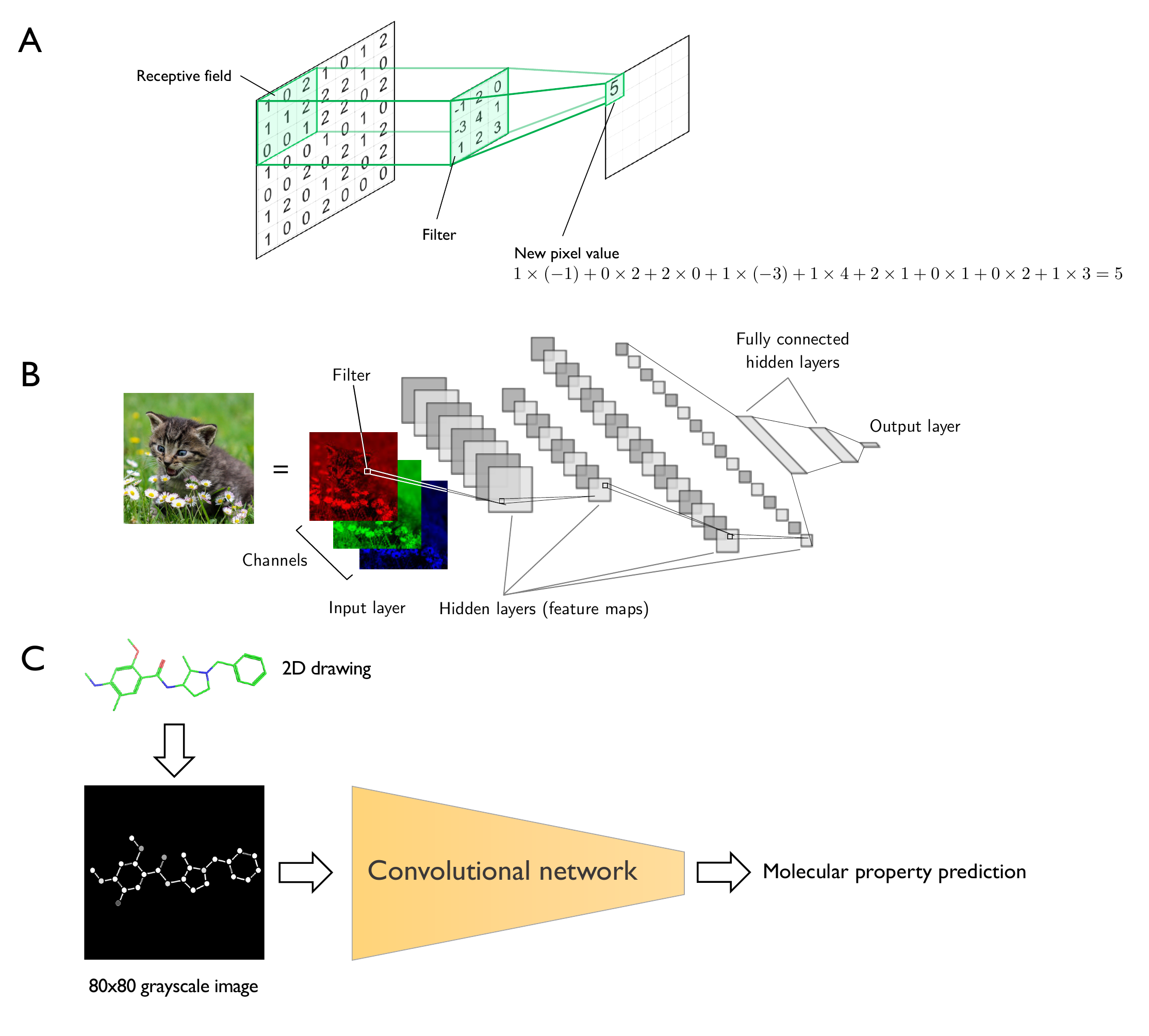}
\caption{{(A) Illustration of a single step in a 2D convolution on an 8x8 pixel input image, using a filter of size 3x3. At each stage, the convolution operation performs a pairwise multiplication of the values in the receptive field and filter. The pairwise multiplication results are then summed to obtain the output pixel. For the next step, the receptive field is moved to the next position (not shown), to compute the next output pixel. This process is repeated until all output pixels are computed.} (B) Depiction of a convolutional neural network (\ac{CNN}) architecture for image recognition, which consists of multiple feature maps that are connected via convolutional layers. Each unit in a feature map is connected with a local patch of the previous layer's feature map through a set of filters. Traditionally, pooling layers have been inserted between every other feature map but are not necessary~\cite{springenberg2014striving}. The fully connected hidden layers and output layer resemble an MLP architecture. {(C) Application of a CNN for molecular property prediction as described in~\cite{goh2017chemception}.}}
\label{fig:cnn-1}
\end{figure}

Convolutional neural networks (CNNs) are feedforward neural networks that utilize the discrete convolution operation as a filtering operation on multi-dimensional arrays (Figure~\ref{fig:cnn-1} A). Depending on the architectural design and discrete convolution operation, CNNs can handle various data modalities, including 1D (sequences and signals in vector form), 2D (pictures and audio spectrograms), and 3D arrays (such as videos or images with depth dimension like 3D computerized tomography scans).  

Today, CNNs are the most widely used deep learning architectures. They are widely recognized for their state-of-the-art performance for image-related tasks such as object classification, detection, and segmentation. The first successful demonstration of CNNs for image classification was the design of the LeNet architecture for handwritten digit and document recognition in the 1990s~\cite{lecun1990handwritten,lecun1998gradient}. While CNNs have shown exceptional performance on various image recognition tasks, including the segmentation of biological images~\cite{ning2005toward,lecun2015deep}, the rise of CNNs to mainstream success can be attributed to 2012 ImageNet competition, where CNNs were able to outperform traditional computer vision methods by doubling the object classification accuracy on a database consisting of millions of images~\cite{krizhevsky2012imagenet}.  

Similar to MLP architectures, CNNs are feedforward neural networks. {However, in contrast to MLPs, CNNs rely on convolutional rather than fully connected layers at the core of the architecture.} The main concepts behind CNNs that distinguish this architecture from MLPs are {\it sparse connectivity} and {\it parameter sharing}. Here, sparse connectivity refers to the fact that each unit in the hidden layer is only connected to a small number of units in the previous layer (Figure~\ref{fig:cnn-1} B). The connection occurs through a filter (Figure~\ref{fig:cnn-1} A), which is a parameter matrix that is commonly interpreted as a feature extractor. Via the filter, information between neighboring pixels is combined to compute the activation of a unit in the next hidden layer. This allows CNNs to compose hierarchies of local features to recognize complex shapes in later layers. For instance, in an image classification task, a multi-layer CNN may learn to detect simple geometric shapes like edges and curves in the earlier layers, and in later layers, it can identify more sophisticated features such as cars or buildings based on the underlying geometric shapes. These feature detectors typically only cover a small portion of the input image or feature map, and in the process of the convolution operation, they can be thought of as a sliding window over the image. Hence, the weights for the different patches of the input image or feature maps are shared within a given layer, which is inspired by the fact that a feature detector that works well in one region may also work well in another part. The advantage of weight sharing is that it reduces the number of parameters that need to be learned, which improves computational efficiency and can reduce overfitting, compared to fully connected networks.  

In traditional CNN architectures, feature maps were followed by pooling layers~\cite{lecun1990handwritten,lecun1998gradient}. These pooling layers combine similar features from neighboring regions in the previous feature map into a single feature, which is supposed to help neural networks to learn more complex relationships or geometrical shapes. However, it has been shown that pooling layers are not a requirement in modern CNN architectures~\cite{springenberg2014striving}.  

Deep learning is a fast-moving field, and even revolutionary architectures such as AlexNet, which outperformed state-of-the-art computer vision methods at the ImageNet competition in 2012~\cite{krizhevsky2012imagenet}, have long since been succeeded by other reference architectures. For example, the VGG architecture has achieved notable performance by shrinking the receptive field and stacking many more layers than AlexNet~\cite{simonyan2014very}. The ResNet architecture introduced residual connections, which are shortcut connections between layers that help with vanishing gradient problems and thus improve the training of very deep neural networks~\cite{he2016deep}. Inception networks improved the extraction aspect of both local and global features by using multiple filter modules in parallel~\cite{szegedy2016rethinking}. While the design of CNN architectures still remains an active area of research, recent efforts have been targeted towards improving efficiency, for example, MobileNet~\cite{howard2017mobilenets} and EfficientNet~\cite{tan2019efficientnet}. In a benchmark study, Canziani et al. compare both the predictive and computational performance of these models for further reference~\cite{canziani2017analysis}.  

{Goh et al. described a simple application of 2D convolutional networks for molecular property prediction.~\cite{goh2017chemception}. Here, the researchers trained a convolutional network, called Chemception, to predict HIV activity and toxicity (as binary classification tasks, "active" or "inactive"), and solvation (regression) from 2D molecule drawings (Figure~\ref{fig:cnn-1}). To prepare the input data for the Chemception model, the researchers converted 2D drawings of molecules into $80\times80$ pixel grayscale images. In this representation, atoms were represented as dots with different shades of gray to encode atom type information. Similarly, bonds were encoded in two different shades of gray to distinguish between single and double bonds. The images were randomly rotated during training to make the network robust towards different ways to orient a molecule. The authors noted that Chemception slightly outperforms molecular fingerprint-based approaches in activity and solvation prediction but slightly underperforms in toxicity prediction. State-of-the-art convolutional network approaches, including 3D voxelation, are discussed in Section~\ref{sec:3d-voxels}.}

\subsubsection{Recurrent neural networks} 

\label{sec:rnns} 

\begin{figure}
\centering
\includegraphics[width=0.65\textwidth]{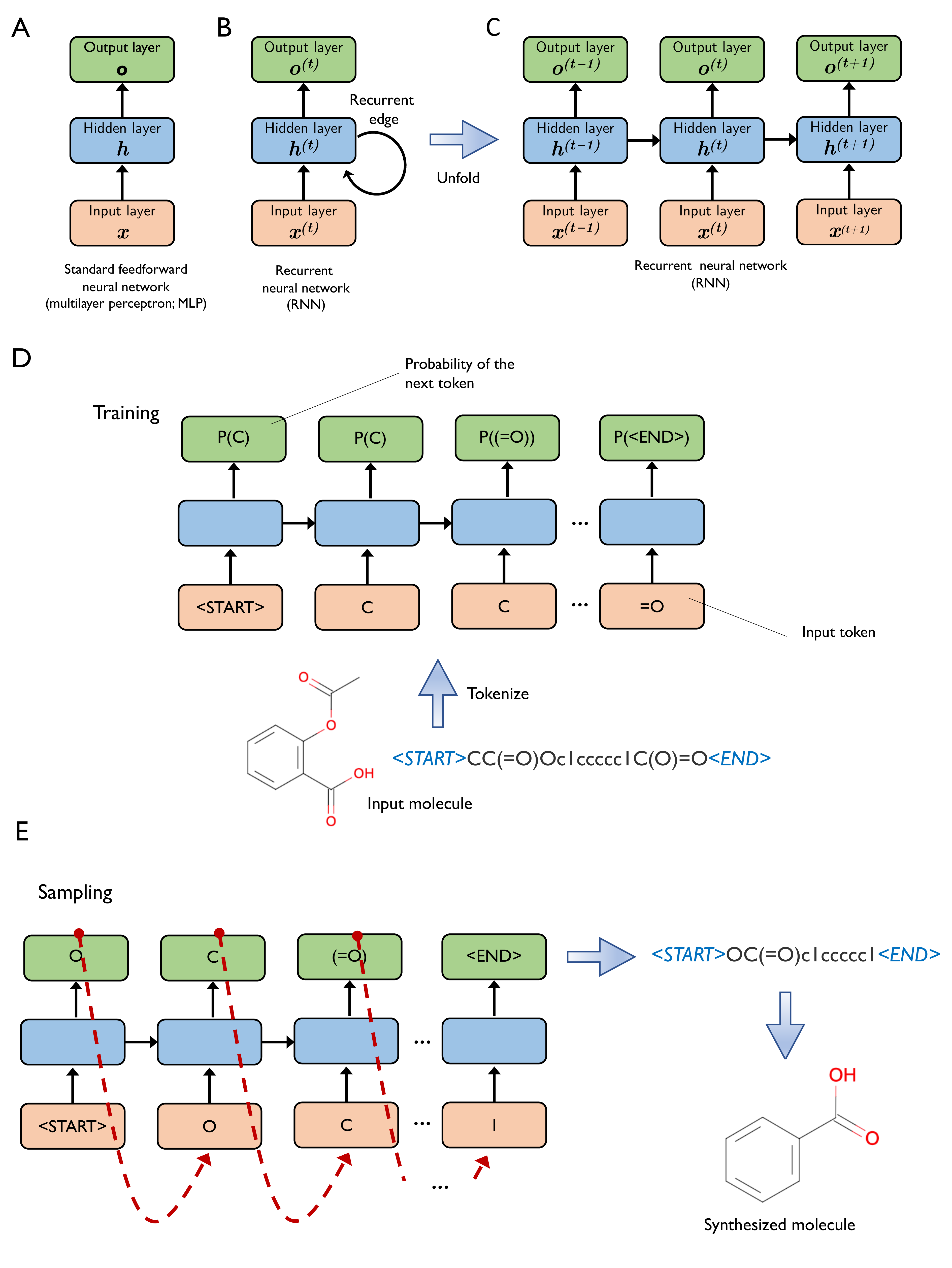}
\caption{Subpanels (A) and (B) depict an MLP with 1 hidden layer and an RNN with 1 hidden layer side by side for comparison. Here, the input layer $x$ containing the input features, the hidden layer ${h}$, and the output layer ${o}$ represent vectors with multiple units. Subpanel (C) depicts the RNN in its unfolded representation. In contrast to an MLP, the hidden layer of an RNN receives its inputs from both the input layer and the hidden layer of the previous time step, $t-1$. {(D) Illustration of the RNN training process. Input molecules are converted into a sequence representation (here: SMILES string), which is divided into tokens. At each time step, the RNN receives one token and returns a probability distribution over all possible token types. The RNN is trained such that the probability corresponding to the next token in the sequence is maximized. (E) To synthesize new molecules after training, a token is randomly sampled from the probability distribution at each time step (tokens with a higher probability are more likely to be drawn). The sampled token is then provided as input at the next time step. When the network outputs an <END> token, the sampling is terminated. The generated SMILES string can then be used as a template for synthesizing the new molecule.}}
\label{fig:rnn}
\end{figure}

Recurrent neural networks (\ac{RNN}s) were originally designed for tasks that involve sequence data, for example, textual data and audio signals such as speech. While traditional RNNs consist of fully connected layers similar to MLPs, RNNs have recurrent connections through time. This recurrence property allows RNNs to maintain an internal history of sequence elements that it processed previously (Figure~\ref{fig:rnn}).

While RNNs can model arbitrary long sequences, the recurrent edges make this architecture particularly prone to vanishing and exploding gradient problems~\cite{pascanu2013difficulty}. One solution to this problem is the so-called long short-term memory mechanism (LSTM) by Hochreiter and Schmidhuber~\cite{hochreiter1997long}, which is a memory cell that replaces the hidden layer of conventional RNNs. A mechanism similar to LSTM, although slightly more computationally efficient, is the Gated Recurrent Unit (\ac{GRU})~\cite{jozefowicz2015empirical}. 

{Merk et al.~\cite{merk2018novo} described an RNN model for designing new drug-like molecules with desired properties. The RNN was first trained on a large dataset consisting of SMILES strings of known bioactive molecules. Then, the researchers used transfer learning Section~\ref{sec:transfer-learning} to fine-tune the model on retinoid X and peroxisome proliferator-activated receptor agonists to generate target-specific ligands. The general workflow is summarized in Figure~\ref{fig:rnn} C-E.}

The remainder of this paper is organized as follows. First, we discuss one of the common challenges with applying machine learning to bioactive ligand discovery, namely choosing an appropriate feature representation for the ligand structure, and highlight recently developed deep learning methods that use these representations. Next, we discuss the latest trends for ligand-based analysis, from molecular property prediction to similarity-based virtual screening.  While the ligand-based approaches assume that a high-quality receptor structure is not available, the next section reviews the recent developments for receptor structure-based bioactive ligand discovery.  {We then explore advances in {\it de novo} small-molecule design.} Finally, this review concludes by motivating the use of transfer learning, which allow researchers to make better use of publicly available data in machine learning and AI-based bioactive ligand discovery.

\section{Molecular feature representations}

\label{sec:molecular-representation} 

Machine learning methods excel at prediction tasks across multiple disciplines but require careful data preparation as most methods are designed to operate on tabular datasets. The standard data input format is the so-called \textit{design matrix}, where each row represents a new training example, and the columns correspond to the different feature variables, {as illustrated in the example in Figure~\ref{fig:supervised-workflow}}. A common challenge in conventional machine learning is how to prepare datasets as input to machine learning algorithms -- in practice, machine learning practitioners have to find a sweet spot between reducing the dimensionality and retaining salient information that the model can learn from. In contrast to conventional machine learning, deep learning excels at learning from \textit{raw} data, such as images and text, directly, {as previously discussed in Section~\ref{sec:dl-architectures}}. However, molecular data, such as conformations of small molecules and receptors, can be challenging to represent in a standard format that most machine learning and deep learning methods have been designed for. Even if the same information can be extracted from two different data representations, an algorithm may be more effective at extracting that information from one over the other. There is no clear best representation of molecules for machine learning methods and indeed certain representations may be better for certain tasks. The following section provides a brief overview of commonly used molecular representations as well as some recent applications of them using AI-based methods.

\subsection{Property-based feature vectors}
\label{sec:property-vec}

A molecular descriptor is the transformation of chemical information into a numeric value~\cite{todeschini2009molecular}. Dragon~\cite{mauri2006dragon} and Mordred descriptors~\cite{moriwaki2018mordred} are examples of sets of molecular descriptors. As an alternative to molecular descriptors, molecular fingerprints encode molecular structure in a vector format, a so-called bit vector consisting of 1's and 0's. When used as input for machine learning models, both molecular fingerprints and descriptors have historically produced state-of-the-art results on chemical machine learning tasks such as chemical odor prediction and bioactivity~\cite{keller2017predicting, ballester2010machine}.

The extended connectivity fingerprint (\ac{ECFP}) is among the most widely-used 2D fingerprint methods~\cite{rogers2010extended}, and we use its generation procedure as an example of the general process for generating traditional molecular fingerprints. A fingerprint is generated by a multistep process in which each atom is associated with a series of integers. In this series the $k$th integer encodes information about the atom it is associated with as well as information about the atoms and bonds within $k$ bonds of that atom -- that is, the substructure of the compound that is within $k$ bonds of the atom. Next, the integers associated with each atom are concatenated into an array format, which is then processed via a hashing algorithm to generate a bit vector of a desired length (typically 1024 or 2048 elements). This method captures information about all identified substructures in a compound, resulting in a fixed-length vector regardless of the input compound's size. ECFPs do not explicitly encode the 3D spatial information of a compound; however, specialized fingerprint methods have recently been developed that incorporate 3D-strucutral information~\cite{axen2017simple}. Lastly, there are also fingerprints that can encode protein-ligand interactions~\cite{da2014structural}.  

\subsection{SMILES}

Simplified molecular-input line-entry system (SMILES) strings are ASCII string representations of compounds (Figure ~\ref{fig:representation-figure} A), which are generated according to a procedure that guarantees a unique mapping from a SMILES string to a compound structure (though not the inverse)~\cite{Weininger1988smiles}. One benefit of SMILES strings over 2D molecular fingerprints like ECFP is that they encode stereochemistry explicitly. One downside for machine learning is that SMILES do not have a fixed-length; however, certain deep learning architectures designed for processing text documents, like RNNs or 1D CNNs, can handle variable-length inputs. 

Recently, Hirohara et al.~\cite{hirohara2018convolutional} proposed a novel molecular representation scheme by converting SMILES strings into "SMILES feature matrices," which were used as inputs into a 1D CNN~\cite{hirohara2018convolutional}. A SMILES feature matrix was constructed by mapping a SMILES string of length $N$ to a $N \times 42$ matrix, where the $k$th row represents the $k$th character (corresponding to either atom or connectivity information) of the string, and the 42 columns correspond to properties of that character. To address the problem of varying-length input strings, the feature matrix was padded with rows of zeros such that all feature matrices had their number of rows set to the length of the longest SMILE string. The predictive performance of the resulting CNN model was comparable with other deep learning methods on the Tox 21 dataset, which contains 8,000 compounds labeled as active or inactive for 12 proteins~\cite{huang2016tox21challenge}.  More interestingly though, this novel feature representation method enabled the extraction of a 64-dimensional vector from a convolutional layer referred to as the \textit{SMILES convolution fingerprint}, which can be mapped back to the model input SMILES, providing an interpretable data-driven fingerprint. 

Another notable recent development is the SMILES2vec method, which combines components from both recurrent and convolutional neural networks~\cite{Goh2017smiles2vec}. Here, the model input is a SMILES string converted into a one-hot encoding of characters present in the SMILES in the training dataset. The one-hot encoding then serves as an input to a 1D convolutional layer, which is followed by two GRUs before the fully-connected output layer that returns the target value. This architecture achieved equivalent performance to the state-of-the-art at the time on the ESOL solubility dataset~\cite{delaney2004esol}.  

\subsection{3D voxels}
\label{sec:3d-voxels}

Molecular structures can be considered as 3D objects, and modeling them as such would encode their structural properties effectively. The conventional way for encoding 3D objects is by voxelization, which, in this case, takes a 3D space and discretizes it into a 3D grid. Each unit cube of the 3D grid represents a voxel (Figure ~\ref{fig:representation-figure} C), which can be considered as the equivalent of a 2D pixel in a 3D space. Instead of merely labeling each voxel via a binary membership indicator based on what part of the compound occupies the voxel, a feature vector with discrete or continuous-valued attributes can be assigned to each voxel, which can provide further information about the atom type or charge, for example. This adds an additional dimension to the representation.  Since this representation is typically associated with a large number of input features, depending on the resolution of the voxel space as well as the size of the vector representation at each voxel position, deep learning approaches used with this type of input representation typically rely on 3D convolutions. This is equivalent to using 2D convolutions, which are commonly used for image analysis (Section~\ref{sec:cnns}), in the 3D voxel space. Unfortunately, these convolutions can still be prohibitively slow, even at relatively low voxel resolutions, which can result in a coarse representation of the compounds properties. Since each cube represents a unit of measurement, and compounds can vary in size, this representation needs to be padded so that it can accommodate for the largest compound in the dataset, which exacerbates the computational efficiency challenges of this method. Likely owed to being computationally very intensive and inefficient, this method is not a common choice for ligand-based virtual screening. However, this representation has shown to be successful for structure-based VS where binding pocket size is consistent regardless of the interacting ligand, and capturing the 3D structure of the protein-ligand complex is important for the task. 

We examine the 3D voxelization process from Ragoza et al. as a concrete example~\cite{ragoza2017protein}. In this publication the authors create 3D voxel representations of high-scoring docking poses of protein-ligand complexes for use in a 3D CNN that computes binding affinities. This representation voxelizes a 24  {\AA} cube centered on the docked ligand into .5  {\AA} voxels. Each voxel has channels for Smina atom types in the ligand and separately for Smina atom types in the protein~\cite{koes2013lessons}. The value contributed to a channel by an atom of the type that is associated with that channel is based on a function of two values: First, the distance between the center of the voxel and the center of that atom, and second, the atom's van der Waals radius." This function is a continuous piece-wise combination of a Gaussian and a quadratic based on these two values.

\subsection{Graph representation}
\label{sec:graph-rep}
Molecules are commonly visualized as undirected graphs. In this representation, atoms represent the graph's nodes, and the bonds are the graph's edges. A naive approach that utilizes such graphical information is to consider pictures of molecular graphs as inputs to a DNN. This has been tried in the literature~\cite{goh2017chemception} using 2D CNN architectures that have been effective for image classification~\cite{goh2017chemception,Meyer2019learning}. However, these models do not outperform MLP and random forest trained on ECFPs. Two issues with this approach are that atomic properties and spatial relationships must be inferred implicitly and that the representation is sparse, with lots of white space that provides little chemically relevant information to the CNNs. Both these issues can be addressed with graph neural networks (GNNs)~\cite{duvenaud2015convolutional,gilmer2017neural}. 

\begin{figure}
\centering
\includegraphics[width=0.8\textwidth]{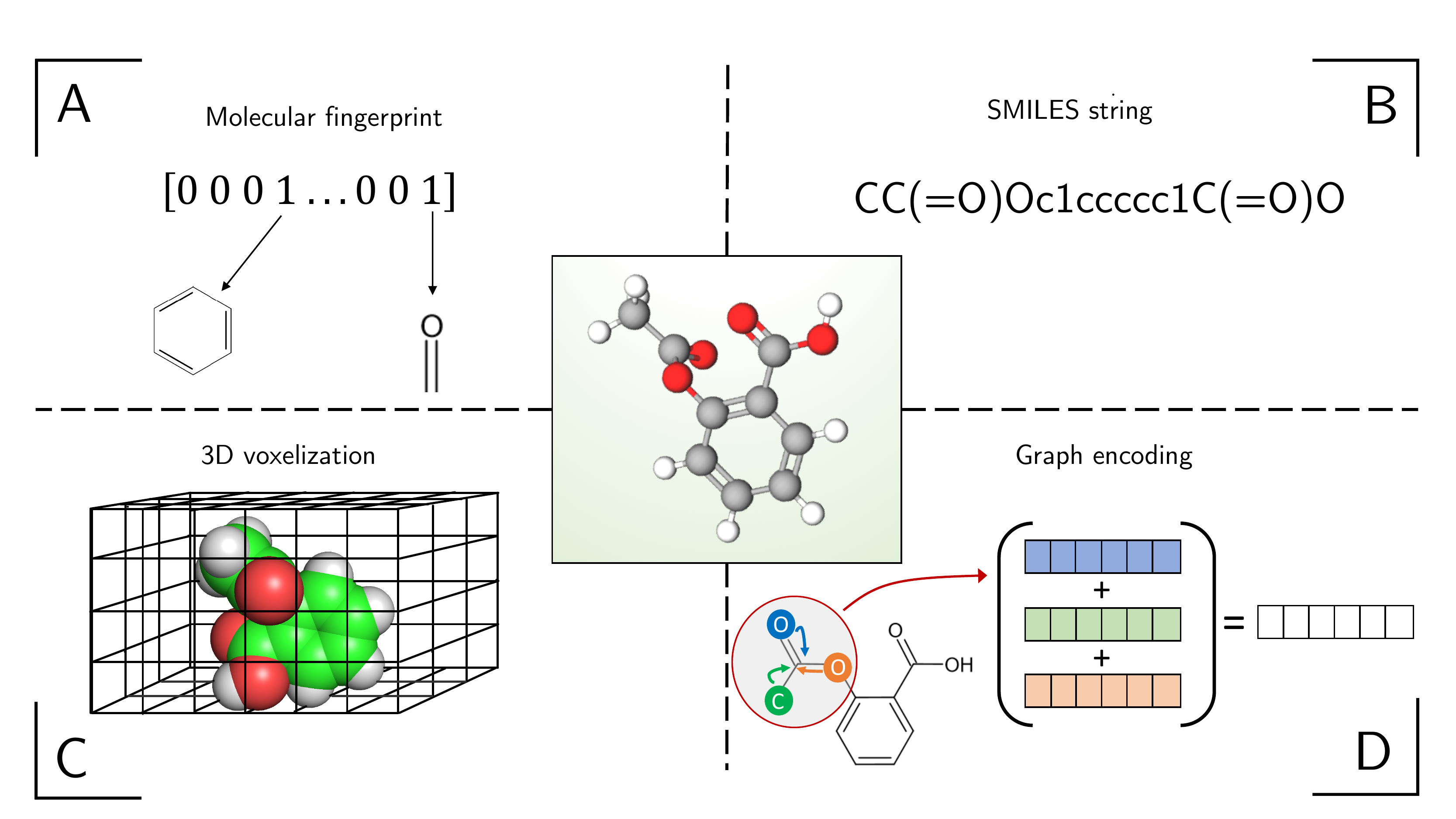}
\caption{Summary of commonly used molecular representation methods based on the example of Aspirin (shown in the center). (A) A molecular fingerprint encodes structural motifs into a sparse bit vector. (B) A SMILES string encoding structural information of the molecule as well as its stereochemistry. (C) A visualization of the  3D voxelization concept. Note that information about which atoms occupy which voxels would be encoded in a 4th dimension which is omitted in this visualization. (D) Illustration of how information is passed to an atom in a simple graph neural network. Note that the graph-structural information will be passed from more distant atoms when the summation is repeated (not shown).}
\label{fig:representation-figure}
\end{figure}

NNs operate directly on a molecular graph (Figure~\ref{fig:representation-figure} D). Similar to images, graphs can encode local structure, but the local structural information is based on the graph structure rather than Euclidean distance. Like CNNs, GNNs utilize sparse connectivity and parameter sharing, but the connectivity is based on the graph's structure. For a detailed explanation of graph convolutional operators, we refer to~\cite{gilmer2017neural}. On a high-level, graph convolutions are based on message passing frameworks for GNNs. At each node of the graph, the following steps are performed: a {\it message} function is applied to each of a nodes' neighbors individually, and the outputs are then added. An {\it update} function is then applied to the summed message functions and the current node with the output being the updated value for the current node. After graph convolutions are performed, the information in the graph can be aggregated by a {\it readout} function, which produces the networks' prediction output.

{After choosing an input representation, the training process and evaluation procedure of machine learning models tend to follow a general workflow similar to the illustration in Figure~\ref{fig:supervised-workflow}. While many state-of-the-art machine learning methods are introduced in the context of general molecular benchmark datasets, most applications of these models can be mapped to this general workflow and adopted for GPCR bioactive ligand discovery.} {For many machine learning methods discussed in the following sections, and especially for DNNs, there are numerous training parameters to tune and many approaches for doing so. We do not focus on these approaches but refer to the respective publications that discuss them further. Similarly, there are many approaches to evaluating models that are beyond the scope of this review. Instead, the focus of this article is on the different workflows enabling bioactive ligand discovery and advances in GPCR machine learning models to help internalize the high-level approach to research in this field.}

\section{Ligand-based methods}

Ligand-based \ac{VS} methods are traditionally defined as methods that only rely on physicochemical and structural information about the ligand and sometimes measurements of ligand-receptor interactions. No other information about the target, like the receptor structure, is utilized.  The use of ligand-based methods is particularly appealing when working with GPCRs for which only a small set of high-quality structures exist.  

In the following subsection, we review notable advances in predicting molecular properties, which are applicable to GPCR bioactive molecule VS.

\subsection{Molecular property prediction}
\label{sec:molecular-property-prediction}

{In the following subsection, we explore recent work done in predicting molecular properties that has been applied to GPCR-related properties, and explore advances in the space that have been or could be applied to GPCRs. }

\subsubsection{Predicting bioactivity from conventional fingerprint representations}

{Researchers applied a variety of standard machine learning algorithms to the task of classifying compounds as active or inactive with two GPCRs, cannabinoid receptor 1 and cannabinoid receptor 2~\cite{bian2019prediction}.  In addition to the activity prediction for cannabinoid receptor 1, the researchers also predicted if compounds acted in an allosteric or orthosteric fashion. The datasets used in this study were constructed from multiple chemical databases. Orthosteric ligands were selected from ChEMBL~\cite{gaulton2016chembl} based on an inhibitory constant ($K_i$) cutoff. Allosteric ligands were selected from the Allosteric Database~\cite{shen2015asd}. The ZINC database~\cite{sterling2015zinc}, a collection of commercially available chemical compounds, was used to add additional inactives. The authors generated three different vector representations of this ligand data: ECFP fingerprints, MACCS fingerprints, and a set of handpicked molecular descriptors. Stratified subsampling was used to create the training and test datasets; stratified subsampling is a procedure that maintains the ratio of active to inactive, or orthosteric to allosteric, compounds across sets. The data was then used to train and compare the performance of a set of machine learning classifiers on each task with each of the three data representations. The set of machine learning classifiers consisted of an SVM, MLPs with 1-5 hidden layers, a random forest, a na\"ive Bayes classifier, and a logistic regression classifier. There was no clear top-performing model across data representations when the models were evaluated with a variety of performance metrics. However, for individual data representations, an MLP with only one hidden layer performed best overall with the MACCS featurization, while logistic regression performed best in combination with the ECFP featurization. This all serves as an example of the standard machine learning workflow being utilized in GPCR research (Figure~\ref{fig:supervised-workflow}). In addition to evaluating the predictive performance of the different models, the researchers also evaluated the importance of the features in each representation. This was done by recursively removing the features of least importance to a model and retraining models until a user-specified number of features was reached. The researchers then examined how important features differed among ligands with different properties. For example, with MACCS features, which represent molecular substructures, it was found that around 10\% of the allosteric ligands had an amide group attached to an aromatic substructure, while this was only the case for 1\% of orthosteric ligands. This type of analysis could be applied to other GPCRs, to elucidate further what ligand components are important for certain interactions.}

{\subsubsection{Learning fingerprint representations}}

{The previous example uses standard machine learning models that take a vector representation of a ligand as input. This allows feature importance to be assessed in ways that would not be readily possible with DNNs that operate on other types of input representations, such as images or graphs. However, most models that achieve state-of-the-art performance in molecular property prediction incorporate deep learning along with novel ligand representations. One such model is WDL-RF, an innovative architecture for predicting molecular properties that was trained and tested on a suite of GPCR ligand datasets to predict ligand bioactivity~\cite{wu2018wdl}. The novel architecture of the WDL-RF model consists of a DNN that takes a graph representation as input (Section~\ref{sec:graph-rep}) followed by a random forest regressor that takes the DNN output as input. The NN is trained to produce an embedding vector the authors refer to as a "data-driven molecular fingerprint." More specifically, this DNN is a \ac{GNN} (Section~\ref{sec:graph-rep}) that shares weights in the network based on the number of bonds attached to each atom node. Each atom's initial feature vector includes a one-hot encoding of its element, the total number of atoms connected to it, the number of attached hydrogen atoms, its implicit valence, and an aromaticity indicator. After a series of graph convolutions, the GNN aggregates the graph embedding into a vector, the data-driven molecular fingerprint, which is used as input for the random forest that is trained to predict the molecular property of interest.}

{The performance of WDL-RF was evaluated on twenty-six GPCRs from families A, B, C, and F. For each GPCR, at least two hundred known ligands were collected from the Uniprot and GLASS databases~\cite{uniprot2019uniprot,chan2015glass}. The bioactivity data was obtained from ChEMBL~\cite{gaulton2016chembl}. In addition, the authors also included known GPCR ligands from ChEMBL that do not interact with the 26 GPCRs evaluated in this study to improve the model's robustness. This aspect of their workflow is worth highlighting because when applying machine learning to GPCRs in general, data quantity is often a limiting factor for using more powerful methods. After splitting the dataset into training and test datasets for each GPCR, the WDL-RF was trained in two steps. First, a GNN was trained to predict ligand bioactivity. Secondly, the fingerprint embedding produced by the GNN was used to train a random forest regressor to predict bioactivity. The second step was necessary to allow fair comparisons between WDL's data-driven fingerprint representation and other popular molecular fingerprints since random forests operate on feature vectors rather than graphs. The WDL-RF model, evaluated via the root mean squared error between predicted and actual bioactivity measures, significantly outperformed hand-crafted molecular fingerprints like ECFP and MACCS on the majority of GPCRs. Additionally, WDL-RF consistently outperformed other neural network-based fingerprints~\cite{duvenaud2015convolutional}. Notably, the few cases where a traditional molecular fingerprint outperformed WDL-RF corresponded to GPCRs for which only a small number of active ligands were available, close to the minimum 200 required for inclusion, which suggests that the datasets were too small to train DNNs effectively.}  

{\subsubsection{Predicting molecular properties beyond bioactivity}}

GNNs have also found success in predicting other molecular properties beyond bioactivity. For instance, they have achieved state-of-the-art performance in odor descriptor prediction~\cite{sanchez2019machine}. Odor perception involves 300 to 4000 different types of olfactory receptors, which are rhodopsin-like (class A) GPCRs~\cite{su2009olfactory}. The odor prediction GNN used {\it atom-node} vectors based on atom type, charge, etc. One unique aspect of this model was that it was trained to predict 138 odor descriptors (fruity, stinky, etc.) for each molecule simultaneously, as the authors found that the correlation structure between certain descriptors helped model performance. A similar simultaneous prediction approach could be taken with other molecular properties that are relevant to GPCRs if they are believed to have correlation structure.  Similar to WDL-RF, when an embedding within the GNN was used to train a random forest model it was found to outperform traditional fingerprints and molecular descriptors.

{The above result suggests that different compound representations capture chemical information in ways that are utilized more or less effectively by various machine learning algorithms. If a model takes multiple representations as input, the ensemble may have better performance. The model CheMixNet followed this approach and used both a SMILES and fingerprint representation of a molecule as input~\cite{paul2018chemixnet}. The authors considered four subnetworks, three of which received a one-hot SMILES string as input and one that was trained on a MACCS fingerprint representation. Each of the three SMILES subnetworks consisted of GRU cells, 1D convolutional layers, and a combination of 1D convolutional layers and GRU cells; the fingerprint subnetwork was an MLP. The researchers trained models with combinations of these subnetworks, where the outputs of each subnetwork were concatenated and passed through additional fully connected layers before a linear or softmax layer to predict the property. The model was trained and tested multiple times to predict a variety of molecular properties. In all cases, certain subnetwork permutations outperformed or matched other state-of-the-art methods at the time. Interestingly, the best subnetwork ensemble for each dataset was not consistent, suggesting that different representations may be better at predicting different chemical properties.}

{\subsubsection{Leveraging unsupervised data for fingerprint learning}}

One issue with the data-driven fingerprints discussed thus far is that they are generated with labeled training data, which restricts the amount of data that can be used to construct a robust fingerprint. The Seq2Seq model attempted to address this issue by learning a fingerprint from SMILES in an unsupervised manner~\cite{xu2017seq2seq}. This model was an autoencoder that used a GRU-based RNN to encode SMILES into a fixed-length embedding vector and then recover the original SMILES from the embedding. After the model was trained, the embedding vectors for ligands in a labeled dataset were generated to use as data-driven fingerprints in other machine learning models. Experiments showed that random forest, gradient boosting, and SVMs that were trained on the Seq2Seq fingerprints outperformed the same models trained on ECFP and the original neural fingerprint produced by Duvenaud et al.~\cite{duvenaud2015convolutional}. One additional benefit of fingerprints generated by Seq2Seq is that they are invertible, so it is easy to recover the original SMILES string if given the fingerprint. 

Shortly after the publication of the Seq2Seq, researchers developed a variational autoencoder (\ac{VAE}) for encoding SMILES into a continuous representation in the latent space~\cite{gomez2018automatic}. This method also produces invertible fingerprints, and a regression model can be trained on these embeddings to predict various molecular properties. The use of a VAE is the game-changer here, as molecules with similar properties are encouraged to be closely together in latent space. As a result, slightly perturbing the embedding of a ligand and inverting it will tend to give a different but similar ligand. This property is of interest in {\it de novo} synthesis which is discussed in Section~\ref{sec:denovo}.   

\subsection{Similarity-based virtual screening}

Similarity-based screening methods are based on the hypothesis that active compounds share similar properties~\cite{johnson1990concepts}. However, there is no trivial definition of compound similarity as computational methods work with approximate representations. For this reason, similarity-based methods heavily depend on and are defined by the molecular representations they operate on. Most machine learning-based methods for ligand-based VS are clustering algorithms, and recent advances in this research are focused on developing new data representation methods and similarity measures. {Several examples of GPCR-specific VS in the literature~\cite{vogt2008exploring,luo2016comparative}. However, recent methods were not specifically on GPCRs but rather evaluated on common benchmark datasets that include several GPCR targets. Nonetheless, their general nature could allow researchers to utilize these methods in GPCR-specific contexts in future research endeavors.} 

\begin{figure}
\centering
\includegraphics[width=0.85\textwidth]{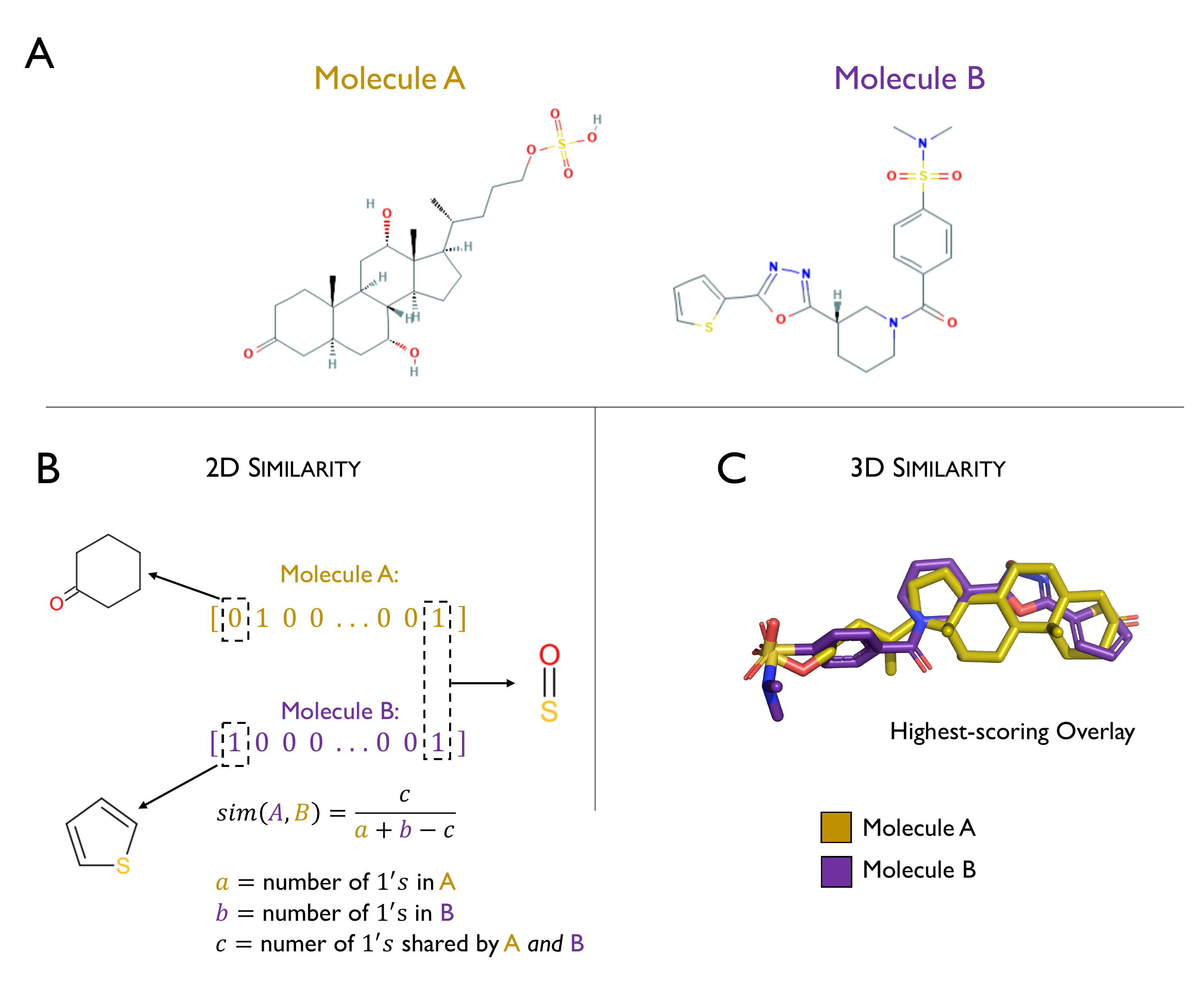}
\caption{{Illustration of molecular similarity between two molecules (A), 3- keto-petromyzonol sulfate (molecule A) and a benzenesulfonamide (molecule B, ZINC ID ZINC38899649). (B) 2D similarity computed by comparing fingerprint representation of the molecules and computing the Tanimoto coefficient as a similarity measure between the two. (C) 3D volumetric overlay, which can be created by software such as ROCS~\cite{hawkins2007comparison}, for which the Tanimoto similarity can be computed from the shapes based on positional densities.}}
\label{fig:2D-3D-sim}
\end{figure}

{\subsubsection{2D similarity measures}}
Tanimoto similarity is commonly used when comparing molecular fingerprint vectors (Figure~\ref{fig:2D-3D-sim} B), as it compares favorably to other similarity metrics on molecular fingerprints, which has been confirmed in analyses based on the sum of ranking differences and ANOVA analysis~\cite{bajusz2015tani}. Fingerprint similarity-based VS is not a new methodology~\cite{willett2006similarity}.  While there are more complex 3D approaches, these methods are still used to query large databases due to their computational efficiency. Furthermore, the efficiency of fingerprint comparison allows for more complex clustering protocols to be tractable.

A recent example of this is the Tanimoto-based fingerprint similarity method MuSSeL~\cite{alberga2018new}. MuSSeL generates an ensemble of different fingerprints for a database of compounds which were labeled with a receptor as their target class and their binding affinity for the corresponding receptor. Given a query compound's fingerprint ensemble, a molecule is considered a candidate for a given target receptor class only if it meets a minimum Tanimoto similarity cutoff with at least $N_1$ fingerprints in the set of known actives. If this requirement is met, the compound must also pass the following criterion for $N_2$ fingerprints -- where $N_2$ is a number that has to be specified by the user, similar to choosing $N_1$. First, the compound must have at least $k$ nearest neighbors above a new user-specified Tanimoto cutoff, and second, the maximum difference between two of the $k$ neighbors' binding affinities must be below a user-specified threshold. The maximum difference threshold is included to help avoid activity cliffs~\cite{stumpfe2012exploring}. For each fingerprint that meets these conditions, the $k$ neighbors' binding affinities are averaged. Finally, for each query compound, the $k$-neighbor averages are averaged for all the fingerprints in the ensemble that passed the second filter to predict the binding affinity of the query compound. The algorithm had good performance on a calibration set and correctly labeled compounds in a number of case studies with GPCR targets such as adenosine $A_{2A}$ receptor and dopamine receptor D4. {This method could be applied to any GPCR of interest to search for new actives, provided there are enough ligands that have measured activity with said GPCR.} 

{\subsubsection{Comparing 3D representations}}
While many similarity-based methods that utilize them are computationally efficient, one of the issues with common fingerprints like ECFP and MACCS is that they do not explicitly capture 3D information about the compounds. This is problematic in situations where a molecule's 3D structure is valuable for assessing a property, such as the steric fit of a predicted pose docked into the receptor's binding site, so methods for comparing 3D representations of compounds are of practical use. 

One approach for comparing 3D representations of molecules could be to find an overlay with maximum shape overlap (Figure~\ref{fig:2D-3D-sim} C). Raschka et al. recently showed that 3D overlay-based VS can be made computationally feasible on large databases if hypotheses based on prior information about the ligand-receptor interactions can be used as filtering criteria~\cite{raschka2018enabling}. Using this hypothesis-driven approach, combined with machine learning-based molecular feature importance assessments, the researchers discovered a potent GPCR signaling inhibitor by screening more than ten million commercially available molecules~\cite{raschka2018enabling,raschka2018automated}. 

One weakness of using just maximum shape overlap to determine molecular similarity is that it does not utilize electrostatic information. {There are multiple analyses that show that electrostatics play an important role in GPCR binding sites, so 3D methods should incorporate them in some capacity}~\cite{baltoumas2013interactions,javitch1999electrostatic}. Thus, similarity scores utilized in this space are generally based on a combination of structural overlap and the overlap of groups with like properties. Methods like ROCS~\cite{hawkins2007comparison} and WEGA~\cite{yan2013enhancing} represent molecules as overlapping Gaussian spheres with each atom sphere containing information about its location, atom type, and charge. The maximum scoring overlay is then based on both the structural and chemical overlap. ESim is a recent method that computes overlay similarity in a different way~\cite{cleves2019electrostatic}. The model uses a set of "observer points" in 3D space that are a fixed distance from one target molecule. At each observer point spatial, angular and electrostatic values are computed that capture properties of the target and query molecule from that observer point's perspective. The closer the values for each molecule are when summed over all observer points, the higher the similarity score is. 

{To evaluate the abovementioned eSim method, the authors used the DUD-E dataset~\cite{mysinger2012directory}. DUD-E is a standard benchmark set for evaluating molecular conformations and docking predictions that includes 102 targets, five of which are GPCR, as well as 22,886 active compounds and their affinities against the targets. Additionally, each active has fifty decoys, molecules that have similar physio-chemical properties but different topologies than the active. An efficient virtual screening algorithm should be able to select actives while avoiding decoys effectively. The performance was evaluated by choosing an active for a target then computing eSim scores for other molecules in DUD-E in relation to that active. From there, a ROC-Curve could be constructed to assess how robustly the model can distinguish between actives and inactives. Statistical analysis of the AUC-ROC scores showed that eSim significantly outperformed other 3D similarity methods, including mainstays like ROCS and WEGA, on many of the DUD-E targets and was only rarely significantly outperformed by another method. For the five GPCR in the dataset, the eSim had higher AUC-ROC in all cases, with significantly better AUC-ROC performance for three of the five receptors.}

{\subsubsection{Hybrid methods and neural network embeddings}}
Methods that rely on 2D fingerprints or 3D overlays for ligand-based VS utilize different information about the molecular structures -- each method has individual strengths and weaknesses~\cite{hu2012performance}. The HybridSim-VS method combines both 2D fingerprint and 3D-structural information~\cite{Shang2017hybridsim}, and the proposed similarity metric is a combination of the Tanimoto similarity of the 3D shape overlay of a pair of molecules and of its 2D fingerprint. This hybrid metric outperformed MACCS, a 2D fingerprint, and WEGA, a 3D representation, when tested on the DUD-E benchmark dataset~\cite{mysinger2012directory}.   

Section~\ref{sec:molecular-property-prediction} summarized different neural network embeddings that can be used as inputs to machine learning models to predict various molecular properties. The same embeddings can be used for clustering-based similarity search. One example of this in practice is a clustering-based SPiDER search~\cite{reker2014identifying}. SPiDER uses self-organizing maps, a special type of artificial neural network approach for unsupervised learning, to create low-dimensional embeddings of the molecule space based on the CATS topological pharmacophores models and physiochemical small molecule descriptors~\cite{reutlinger2013chemically}. Recently SPiDER was used in tandem with a binding affinity prediction via DEcRyPT to discover celastrol, a cannabinoid receptor 1 and 2 agonist~\cite{rodrigues2018machine, rodrigues2019dissecting}. DEcRyPT  is based on random forest models for regression analysis, which have been trained to predict binding affinities from topological pharmacophore features. In this study, the researchers first used SPiDER to select potential cannabinoid receptor agonists via clustering. The selected molecules were then analyzed using DEcRyPT, which was trained on experimental binding affinity values for selected targets, as a predictive model that was applied to the most confident predictions from SPiDER to obtain binding affinity scores (-log10 values). The agonist celastrol was identified by this approach, and the ligand was experimentally validated by biochemical assays including radiological displacement assays, followed by dynamic light scattering measurements to rule out false-positive readouts. The authors emphasized that common similarity-based searches using conventional ECFP4 Morgan-fingerprints would have disqualified celastrol as a GPCR ligand.

\section{Receptor structure-based methods}

{As seen in the previous section, many methods can be effectively utilized to identify bioactive GPCR ligands without explicitly including the receptor structure. Ligand-based methods are the only option when no receptor structure is available, as is the case for many human GPCRs~\cite{lagerstrom2008structural}. However, the structure is informative if it is available and opens the door to new tasks that can be approached with machine learning. We explore applications of machine learning for two such tasks: binding site prediction and protein-ligand docking.} 

\subsection{Binding site prediction}

One of the most notable achievements in binding site prediction for identifying new targets was Nayal and Honig's random forest model, which was trained on 408 features representing structural, geometric, and physicochemical properties~\cite{nayal2006nature}. Their method was able to identify 1,347 cavities on the surface of 99 diverse proteins, and in 100\% of the cases, a drug has been experimentally found to bind at least one of these surface cavities. 

A more recent method for discovering binding sites in GPCRs is implemented in the COACH-D server~\cite{chan2019new,wu2018coach}, which optionally allows users to provide a ligand structure that is then docked into the predicted binding site of the target protein structure to refine the ligand binding site predictions. The prediction in COACH-D is based on the COACH algorithm, which aggregates binding site predictions from both sequence and structure data based on five other methods and then uses a linear SVM classifier for the final prediction~\cite{yang2013protein}. 

Challenges for applying deep learning-based models to binding site identification are two-fold. First, the dataset of available GPCR structures is relatively small. However, this problem can potentially be addressed by transfer learning, which is discussed in Section~\ref{sec:transfer-learning}. Second, deep learning excels when the models can learn to extract patterns from the raw input data rather than hand-engineered feature descriptors as described in Section~\ref{sec:molecular-property-prediction}. While deep learning models based on 3D voxelization methods~\cite{jimenez2018k, li2019deepatom} and graph convolutions~\cite{duvenaud2015convolutional,gilmer2017neural,schutt2017quantum} have been successful in predicting binding affinities and molecular properties from small molecule structures, representing large molecules such as membrane receptors is more computationally prohibitive.

{A recently developed deep learning method addresses the computational challenges of working with membrane receptors by splitting larger proteins into multiple parts~\cite{kozlovskii2020spatiotemporal}. The model BiteNet is a 3D convolutional network that takes 64 {\AA} cubes voxelized into 1 {\AA} units as input. Each voxel has a set of eleven channels for eleven atom types, and each channel's value is the output of an atomic density function applied to the channel's atom type. The output of BiteNet is an 8x8x8x4 tensor, where the first three dimensions correspond to the coordinates of voxel cells in the original 64 {\AA} grid--the coordinates of 8 {\AA} cubes within the 64 {\AA} cube input. The last dimension's four channels correspond to a probability that the associated 8 {\AA} cube contains a binding site and the cartesian coordinates of the most likely binding site. If a protein is larger than 64 {\AA} in any dimension, then it is broken into overlapping 64 {\AA} grids, which are considered separate data points. BiteNet was trained on the frames of molecular dynamic simulation trajectories of a curated set of protein-ligand complexes from PDB~\cite{berman2000protein}. In a case study on the GPCR A2A, the model was able to correctly identify its known binding sites.}

\subsection{Protein-ligand docking}

Structure-based virtual screening for GPCRs is limited by the number of high-resolution GPCR structures that are available. However, advancements in techniques like Cryo-EM have increased the rate at which these structures are solved, such that many general AI-based methods for protein-ligand docking can be applied to modeling GPCR-ligand recognition in the future.  

Structure-based VS methods need to be able to score protein-ligand complexes to identify favorable candidates. Scoring functions can be grouped into four classes: force field, empirical, knowledge-based, and machine learning-based~\cite{li2019overview}.  Popular docking programs typically use scoring from the first three classes~\cite{trott2010autodock,verdonk2003improved,vilar2008medicinal}. Functions from these three classes are sometimes referred to as classical scoring functions. Classical scoring functions are based on linear combinations of features, which restricts the ways they can utilize structural and interaction data~\cite{li2019classical}. Scoring functions based on machine learning are not restricted in this way. One of the first machine learning scoring functions to outperform classical methods was RF-score, which has been updated multiple times to further improve its performance~\cite{ ballester2010machine, ballester2014does,li2015improving}. All iterations of RF-score use the same algorithm for random forest regression. However, the researchers experimented with different ways of encoding the molecular information in the training dataset to improve the performance of RF-score, which underlines the importance of choosing a good input representation as previously discussed in Section~\ref{sec:molecular-representation}. The feature representation of the protein-ligand complex used in the latest version of the RF-score (v3)~\cite{li2015improving} consists of counts of interacting protein-ligand atom pairs. Two atoms are considered interacting if they are within 10  {\AA} in the structure. For example, a carbon in the protein that is 7  {\AA} away from a nitrogen in the ligand in a binding pose would increment the count. These features are not symmetric; The feature for an atom of type $A$ in the protein and an atom of type $B$ in the ligand is a different feature than the one for an atom of type $B$ in the protein and an atom of type $A$ in the ligand. Additionally, RF-score v3 uses features produced by AutoDock Vina~\cite{trott2010autodock}. When trained and tested to model a VS setting, RF-score outperformed classical scoring functions in various evaluation metrics~\cite{wojcikowski2017performance}. 

Initial attempts at developing NN-based scoring functions in this space struggled with overfitting issues and were outperformed by methods that utilize conventional machine learning algorithms, like RF-score~\cite{sunseri2016d3r}. Recently, however, a set of new NN-based scoring functions have shown state-of-the-art performance~\cite{jimenez2018k,li2019deepatom,zheng_onionnet_2019, stepniewska2018development}. All these networks use novel input representations to capture the structural and chemical information of the protein-ligand complex. 

{For models in this space, PDBbind is a relevant database that was used to train all of the NN-scoring functions we review. As of writing, it contains 17,679 protein-ligand complexes along with their experimentally-measured binding affinity data~\cite{wang2005pdbbind}. While not consisting of exclusively GPCRs, models trained on this set could easily be applied to scoring GPCR-ligand poses even if the following methods do not address them specifically.}   

A number of NN-based scoring functions voxelize the 3D representation of the protein-ligand complex (Section~\ref{sec:3d-voxels}) and use functions that capture structural and chemical information to compute the channels of each voxel~\cite{jimenez2018k,li2019deepatom,stepniewska2018development}. All of these methods are based on 3D CNNs, which can be computationally demanding due to their large number of trainable model parameters. Furthermore, the large number of parameters increases the number of training examples required to effectively train the networks without overfitting, which has been identified as an issue for DNNs in this space~\cite{ragoza2017protein}. However, more recent methods aim to address the overfitting issue by using more modern CNN architectures.  For instance,  DeepAtom~\cite{li2019deepatom} was recently designed based on ShuffleNet v2~\cite{ma2018shufflenet}, and the recent $K_{\text{DEEP}}$~\cite{jimenez2018k} method is based on SqueezeNet~\cite{iandola2016squeezenet}. These architectures were designed to achieve good computational performance while keeping the number of trainable parameters relatively small, which helps with minimizing overfitting when deeper architectures are being trained on smaller datasets. Both DeepAtom and $K_{\text{DEEP}}$ achieved state-of-the-art binding affinity predictions on the \textit{PDBbind v. 2016} core dataset~\cite{wang2005pdbbind}. The success of these networks suggests that advances in other subfields of deep learning could be effective in VS as well and could be a fruitful area of future research. 

Other NN-based scoring functions achieved state-of-the-art performance by including more information in the input representations. OnionNet uses atom pair counts similar to RF-score, but generates them for multiple cutoffs and makes each pair exclusive to the first cutoff it appears in~\cite{zheng_onionnet_2019}.  These values are then converted into a $n \times m$ matrix, where $n$ is the cutoff value and $m$ is the atom pair, and this matrix is used as input for a 2D CNN. The use of a 2D CNN reduces the parameter burden of this model so the area of the protein that can be encoded increases substantially. The previously mentioned 3D CNN models, DeepAtom and $K_{\text{DEEP}}$, can only capture molecular information in a 20  {\AA} cube, while OnionNet counts at least all atom pairs within a 61  {\AA} cube. Long-range electrostatic interactions extend beyond 20  {\AA} and are known to be important in protein-ligand binding. Hence, OnionNet may be more effective for VS targets where these long-range interactions are important for activity prediction~\cite{dagliyan2011structural}. In practice, its performance on the \textit{PDBbind v. 2016} core was comparable to the 3D CNNs mentioned above, though there was no performance comparison conducted on individual targets where this information could be useful. 

\section{{\it De novo} small-molecule design}
\label{sec:denovo}

{\it De novo} molecule design is closely aligned with the goals of VS. While VS attempts to answer questions such as "Does this molecule activate this target?", {\it de novo} design attempts to answer "Can we generate a molecule that can activate this target?". Answering the second question in the affirmative implies that we understand what makes a molecule active against a given target. {\it De novo} molecule design can be tied to the idea of inverse QSAR/QSPR, deriving the structures of all molecules that are active with a target or have a certain property. This is a challenging problem, but despite its difficulty, {\it de novo} synthesis has recently become an active area of research, which is likely due to the recent advances in the development of deep generative models and reinforcement learning.         

\subsection{Generating molecules with variational autoencoders}

\begin{figure}
\centering
\includegraphics[width=0.7\textwidth]{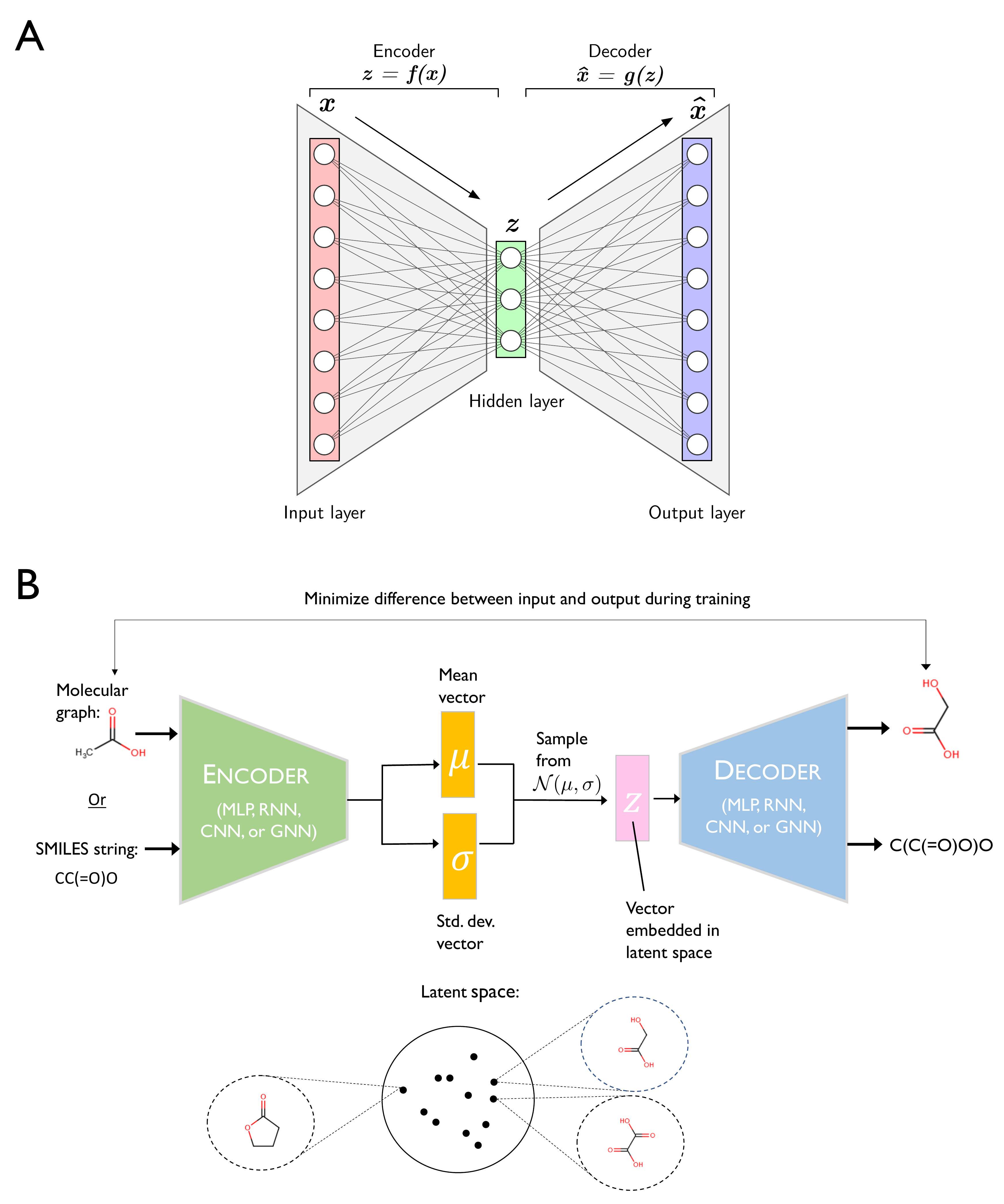}
\caption{{(A) Fully connected autoencoder consists of two submodules: an \textit{encoder} followed by a \textit{decoder}. The encoder submodule compresses an input representation $x$ into a lower-dimensional representation $z$. The decoder converts the lower-dimensional representation to $\hat{x}$, such that reconstruction resembles the original input, $\hat{x} \approx {x}$. (B) Illustration of a variational autoencoder architecture and how it can be applied to molecules. Given a molecular representation as input (acetic acid), an encoder network outputs both a mean vector and a standard deviation vector. These vectors are used as parameters of a normal distribution to sample a vector embedded in the latent space ($z$). This latent vector is used as input for a decoder network, which outputs a molecular representation (glycolic acid). The loss function for variational autoencoder  training consists of two components: the first component is the reconstruction loss ensuring that the original input resembles the generated output; the second component encourages similar molecules to cluster together in the latent space. In this case the structurally similar glycolic acid and oxalic acid are close in latent space, while the less similar gamma-butyrolactone is farther away. After training, the encoder portion can be discarded, and new molecules can be sampled by sampling different $z$ vectors from a standard normal distribution.}}
\label{fig:vae-diagram}
\end{figure}

{Autoencoders refer to architectures that were traditionally used for dimensionality reduction and trained in an unsupervised manner~\cite{hinton2006reducing}. The main idea behind autoencoders is that they learn to preserve the essential part of the data and remove the non-essential ones (Figure~\ref{fig:vae-diagram} A).}

{An autoencoder consists of two neural networks: an encoder that transforms a training example into a lower-dimensional representation, and a decoder that reconstructs the original training example from its lower-dimensional representation. For example, if both the encoder and decoder submodules of the autoencoder are fully-connected NNs with linear activation functions, the latent feature embedding produced by the encoder is equivalent to features transformed via principal component analysis except for the feature orthogonality constraint.}

{It shall be noted that regular autoencoders are deterministic models that can be used for data compression or to modify existing data given a user-specified objective~\cite{mirjalili2018semi,mirjalili2019flowsan,mirjalili2018ensemble,mirjalili2020privacynet}. A related model is the variational autoencoder, which can be considered a generative model as it is capable of synthesizing entirely new, realistic data records mimicking the data contained in the training dataset~\cite{kingma2013auto,doersch2016tutorial}.}

{In essence, a VAE learns to reconstruct high-dimensional inputs from a low-dimensional continuous latent space. In contrast to a conventional autoencoder, points closely together in the latent space are encouraged to have reconstructions with similar properties. A high-level diagram of a molecular VAE is displayed in Figure~\ref{fig:vae-diagram} B. These models take a molecular representation, like a SMILE string~\cite{gomez2018automatic,kusner2017grammar} or a graph~\cite{jin2018junction,simonovsky2018graphvae}, as input. The molecular representations are then passed into an encoder network that outputs a mean vector and a standard deviation vector. These vectors are used as parameters for a multivariate normal distribution, which is sampled from to produce a low-dimensional embedding. The embedding is then passed through a decoder network to get a molecular representation as output. Thus the embedding can be viewed as a lower-dimensional representation of the molecule that is output by the decoder network.}

{The dimensional space that the embedding vector occupies is commonly referred to as the \textit{latent space}. If the goal of an autoencoder is to reproduce its data input, it may seem strange that the encoder network generates parameters to sample data points in latent space instead of generating the data instances directly. However,  this particular design ensures that points close in latent space result in reconstructions with similar properties. Without getting into the details, the parameter vectors, which are output by the encoder network, are used as a regularization term for reconstruction loss of the molecule input into the VAE. This term is minimized when the latent embeddings are sampled from a standard normal distribution. However, if all latent embeddings are sampled from a standard normal distribution, the reconstruction loss will be high.
This encourages the model to group the latent vectors of similar molecules closely together in the embedding space. While this minimizes the regularization term, it also reduces the risk of incurring a high reconstruction loss, because the reconstruction loss for similar molecules will be less than that for different ones. This concept is also conveyed in Figure~\ref{fig:vae-diagram} B -- the molecule generated by the model differs from the input molecule, but they share a relatively large degree of structural similarity and are close in the latent space. This property of the latent space can be utilized to generate novel molecules with similar properties to selected input molecules by taking the input's latent representation and perturbing it in a controlled manner~\cite{gomez2018automatic,dai2018syntax}.  One of the challenges with using VAE-based models in this space is that with current architectures, the VAE has no notion of what constitutes a valid molecule~\cite{jin2018junction}.  Since these models are trained on large numbers of valid compounds, certain models were still able to achieve reasonable results despite producing some invalid structures~\cite{gomez2018automatic,kusner2017grammar}.} 

Jin et al. introduced a novel VAE design that only allowed valid molecules as outputs, called a junction tree VAE~\cite{jin2018junction}. This model uses a GNN to encode a molecule's graph and also a junction tree of valid molecular fragments in the graph. The decoder network first decodes a junction tree from the latent space. Then, it decodes a molecule that satisfies that junction tree. Since all the molecular fragments the junction tree can encode are valid, a valid molecule can always be derived from the junction tree.\cite{wiegerinck2000variational}. In addition to always producing valid molecules, the model outperformed other VAEs on a Bayesian optimization task, where the goal was to find a molecule that maximized octanol-water partition coefficients (logP).

 {One shortcoming of regular VAEs is that the molecular similarity in the latent spaces does not directly translate to the similarity of a molecular property of interest.} As a general example, consider the latent representation of a ligand with a thiol group that induces a positive activity response in a given target receptor. Perturbation of the latent representation of that molecule may result in similar molecules that share the thiol group feature but are not active against the target.  As a consequence, the performance for VAEs in this space is usually assessed on simple chemical properties~\cite{jin2018junction, kusner2017grammar,dai2018syntax}. For this reason, while there is still plenty to be investigated about VAEs for {\it de novo} synthesis, research that has focused on the {\it de novo} generation of compounds for specific targets has largely utilized reinforcement learning, which will be discussed in the next section.

\subsection{Reinforcement learning-based molecular design}

\begin{figure}
\centering
\includegraphics[width=0.6\textwidth]{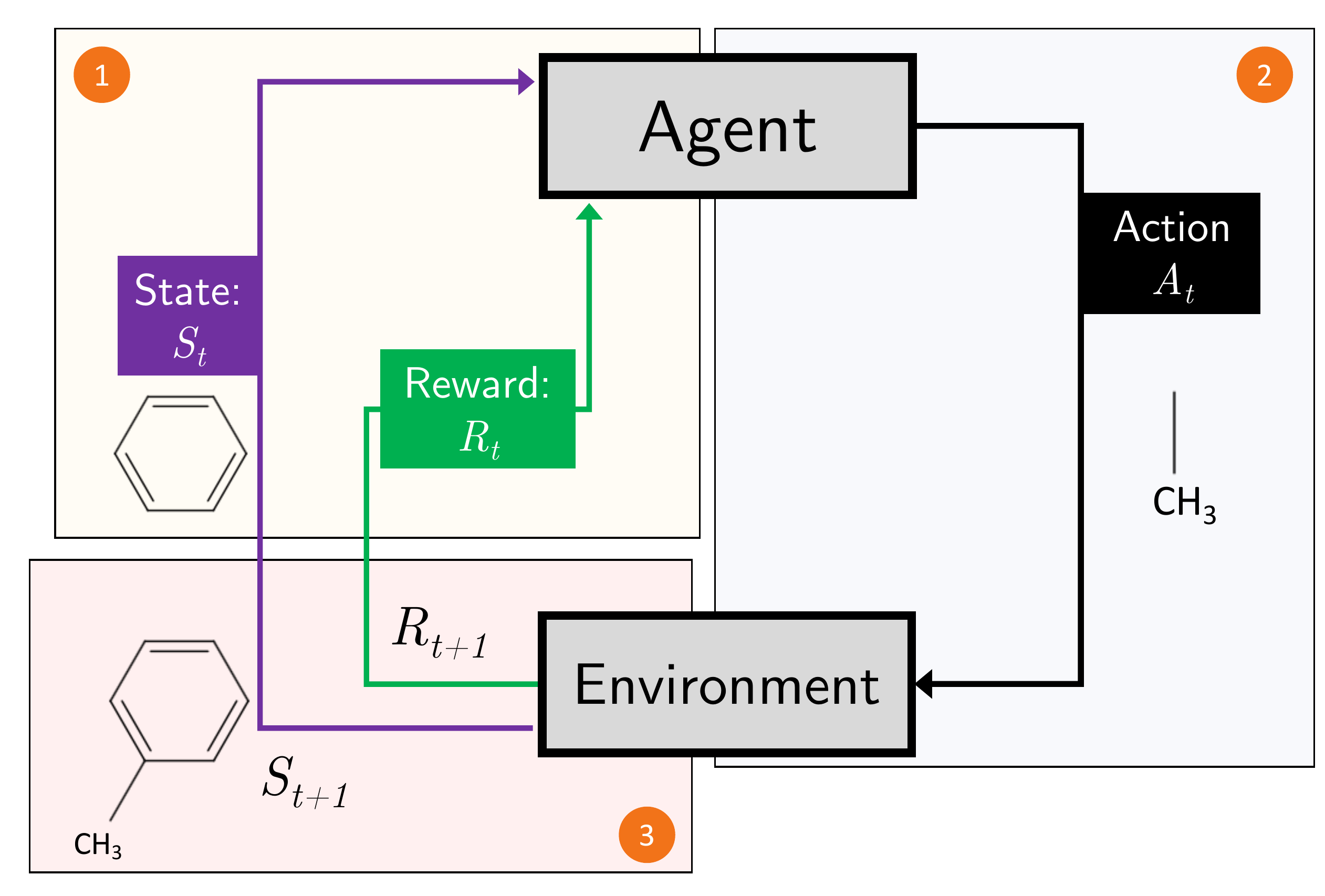}
\caption{Representation of the basic reinforcement learning paradigm with a simple molecular example. (1) Given a benzene ring (state $S_t$ at iteration $t$)  and some reward value $R_t$ at iteration $t$, (2) the agent selects an action $A_t$ that adds a methyl group to the benzene ring. (3) The environment considers this information for producing the next state ($S_{t+1}$) and reward ($R_{t+1}$). This cycle repeats until the episode is terminated.}
\label{fig:rl-diagram}
\end{figure}

Reinforcement learning seeks to teach an agent to perform actions that will maximize a cumulative reward over a series of iterations. In this paradigm, the definition of the reward is very flexible, which makes it possible to more directly optimize for specific traits compared to using  generative models such as VAEs. As illustrated in Figure~\ref{fig:rl-diagram}, at each iteration, the agent is given a state by the environment as well as the reward from the previous iteration. The agent then selects an action based on this information. This selected action is fed into the environment, which then outputs a new state and reward. This cycle continues until the series of iterations, called an episode, is terminated, at which point the agent's behavior is updated based on the information provided by its trajectory, the ordered list of states, actions, and rewards associated with each iteration in the episode. While this high-level overview will suffice to understand how reinforcement learning is utilized for {\it de novo} design, the reader is referred to~\cite{sutton2018reinforcement, raschka2019python}, which we recommend as resources with more in-depth explanations.

 Several examples exist in which reinforcement learning models have been optimized to generate GPCR bioactive molecules.  One such example is the REINVENT model~\cite{olivecrona2017molecular}.  In REINVENT, the agent is an RNN that generates one SMILES string token at a time over the course of multiple iterations. After a batch of 128 SMILES strings are produced, the RNN agent is updated with a modified form of REINFORCE, an algorithm that updates the agent network based on the trajectories of the episodes in the current batch~\cite{williams1992simple}. In one experiment, REINVENT was trained to generate compounds that are active with the GPCR Dopamine receptor $D_2$ (DRD2). In this experiment, the reward function was based on the output of an SVM trained to predict a compound's DRD2 activity. After training, the agent was able to produce compounds that the SVM labeled as active against DRD2 96\% of the time. The reinforcement learning model generated both novel, chemically valid compounds and, impressively, compounds that are known to be active with DRD2 even though these were not included in the training set.     

 DrugEx is another deep reinforcement model that was directly optimized for generating molecules with GPCR activity~\cite{liu2019exploration}. Here, the researchers focused on adenosine $A_{2A}$ receptor, which has been targeted to treat cardiovascular and inflammatory diseases~\cite{chen2013adenosine}. Similar to REINVENT, the DrugEx model also produces SMILES strings over the course of an episode. However, DrugEx adds a stochastic component during training.  Before being used in the reinforcement learning training, the RNN agent network was individually trained with a large set of molecule SMILES from ZINC~\cite{sterling2015zinc}. Two copies of this pre-trained network were then created, with one referred to as the {\it exploration} network and the other as the {\it exploitation} network.  Only the exploitation network was updated during the reinforcement learning training process; however, with a specified probability at each iteration, the exploration network would be queried for the next token instead. The purpose of this procedure was to explore a wider chemical space during training -- afterwards, the exploration network was discarded, and only the exploitation network was used to generate new molecules. This method successfully rediscovered some known actives for adenosine $A_{2A}$ receptor. The authors also suggested the RNN agent was able to produce molecules with a large diversity, by showing the actives generated by the model covered all clusters generated by fingerprint-based clustering on known adenosine $A_{2A}$ receptor actives. 

Other reinforcement learning-based models for {\it de novo} synthesis described in the literature were not specifically focused on GPCRs' bioactive molecule design but could be adopted for such tasks in the future~\cite{zhou2019optimization, you2018graph}. One such example is Zhou et al.'s Molecule Deep Q-Networks (MOLDQN) approach, which modifies deep Q-networks for molecule generation~\cite{zhou2019optimization, mnih2015human}. This model's agent network takes the current molecular graph's Morgan fingerprint as input and selects an action to modify the molecular graph. The actions include adding an atom, adding a bond, or increasing a bond's order. Additionally, the actions that are allowed at a given iteration are restricted if they are invalid so the system will always produce valid molecules. Similar to DrugEx, the model encourages exploration by selecting a random valid action at a given iteration with some probability $\epsilon$. The model achieved state-of-the-art performance when producing molecules that maximized for logP and quantitative estimates of drug likeness separately~\cite{leeson2012drug}.    

While many publications in this area demonstrate the ability to optimize for molecular properties, they unfortunately lack experimental follow-up procedures for further model evaluation. As a notable counter-example, we want to highlight GENTRL, which was used to discover novel inhibitors of discodin domain receptor 1, a tyrosine kinase~\cite{zhavoronkov2019deep}. The inhibitors were generated computationally and then synthesized and experimentally validated. The experiments {\it in silico} and {\it in vitro} were completed in approximately 46 days at a fraction of the cost of a high throughput screening approach.

\section{Transfer learning}
\label{sec:transfer-learning}

Machine learning, and deep learning in particular, requires large training datasets. It's not atypical for modern deep neural networks to have millions of trainable parameters, which require sufficient data for successful parameterization. Transfer learning refers to the process of adapting models that have been trained on one task to another, typically similar task~\cite{pan2009survey}. The most common form of transfer learning is to take a DNN that was trained on a large general-purpose dataset and fine-tune it to a smaller dataset of interest. {As an example, suppose there is only a limited number of ligands with bioactivity data for a certain class A GPCR of interest. However, a larger bioactivity dataset exists for a second GPCR from the same subfamily that has a similar binding site. In this scenario, one could apply transfer learning by training a DNN to predict bioactive ligands for the second GPCR and then fine-tune the model weights on the smaller dataset of the GPCR of interest.}  Typically, the more similar the initial dataset is to the dataset for the target task, the more successful transfer learning can be. {In our example, if the ligands in the larger dataset are physio-chemically similar to those in the smaller one, one could expect transfer learning to be more successful.} 

One example of transfer learning being applied to improve the performance of models with limited available data is in a recent structure-based method that sought to train a 3D CNN to classify molecule activity from high-scoring docking poses~\cite{imrie2018protein}. When the network was trained on targets in a DUD-E based training set and then fine-tuned on a smaller, protein family-specific dataset, the model performed better on test set targets in that protein family than the same network that was trained only on the family-specific set~\cite{mysinger2012directory}. Family-specific models typically outperform general models, but data availability may limit their performance~\cite{ross2013one}. This work shows that transfer learning can help to address this issue. While it has not yet been applied further in bioactivity predictions, transfer learning has been successful in improving the performance of models that predict standard molecular properties like solvency as well as in models that predict quantum mechanical approximations~\cite{goh2018using,smith2019approaching}. 

\section{Future directions}
\label{sec:future-directions}

Currently, one of the most neglected areas of machine learning in bioactive ligand discovery is active learning, which aims to combine artificial and human intelligence. Active learning is a branch of machine learning that focuses on selecting labeled data for supervised learning to improve non-confident predictions and fill the knowledge gaps of the model~\cite{munro2020active}. Knowledge gaps can be filled through human interaction, by sampling optimal unlabeled training examples for annotations by humans. We think that utilizing domain knowledge from experts is crucial in a highly specialized field such as GPCR bioactive ligand discovery, and we believe that it is highly advantageous to develop machine learning systems that include human feedback loops. 

Another avenue that remains largely unexplored for computational ligand discovery is a combination of active learning and transfer learning (Section~\ref{sec:transfer-learning}), called active transfer learning. Here, a separate model is trained on an existing validation set to predict whether a given model is able to make correct predictions or not on a given unlabeled dataset, which then can highlight problematic cases for human review~\cite{munro2020active}. 

Another recent development and promising research trend in deep learning is semi-supervised learning, which is particularly useful if pre-trained models for transfer learning are not available for the target domain or are infeasible to obtain. Semi-supervised learning is the process of deriving and utilizing label information directly from the data itself rather than having humans annotating it. In a language model, for example, semi-supervised learning can be utilized by training a model to predict the next word in a sequence~\cite{howard2018universal} or, in the context of computer vision, this could be the composition of an image into a jigsaw puzzle that the DNN learns to assemble~\cite{noroozi2016unsupervised}. The main idea of this approach is to choose a task that requires an understanding of the underlying data in order to be solved. This stage of self-supervised training can be regarded as pre-training, and the model can then be adopted and fine-tuned to solve the target task downstream similar to conventional transfer learning. 

{In addition to applications of new machine learning paradigms to VS for GPCRs it is also worth considering how machine learning could be used to assist other areas of GPCR research. One potential application for machine learning that could push GPCR research forward is to assist in deorphanization. While this intersection is largely unexplored there is one recent method where researchers used an SVM trained on peptide-descriptor vectors for GPCR neuropeptide ligands to predict if a set of orphaned neuropeptides from \textit{C. elegans} interacted with any putative GPCRs in the organism \cite{shiraishi2019repertoires}. After predicting 22 putative neuropeptide-GPCR pairs, the authors experimentally validated 11 of them with cell-based signaling assays to provide evidence for nine new receptors. A similar approach could be applied to general chemical ligands by modifying the data representation. Such an approach could also be effective in other organisms or with more advanced machine learning models.}

{Binding site modeling is another area that modern machine learning methods could drive forward. As mentioned throughout the paper, one challenge when working with GPCRs is that no high-resolution structures are available. Accurate computational approaches for predicting the structure of a binding site could have an immensely beneficial impact on guiding chemical screening. It is worth mentioning the recently published deep learning model AlphaFold obtained state-of-the-art performance on the protein folding problem, far surpassing traditional approaches to it~\cite{senior2020improved}. While a prediction of binding site structure would ideally require higher resolutions than this model provides, it is a step in the right direction.}

\section{Conclusions}
\label{sec:conclusions}

The confluence of the meteoric rise of deep learning methods, the increase in large chemical datasets, and the further ease of access to high powered computing resources have spurred the field of chemical machine learning forward at a rapid pace. The many representations of chemical data allow a variety of methods used in other fields to be put to task on chemical quandaries and have encouraged the development of new machine learning models. As we have discussed, many of these models have been or could be applied to VS on GPCRs, though the types of GPCR data currently available may limit the general applicability of certain models. However, as the amount of bioactivity data available continues to increase, so will the performance of these models. One area in which the space is currently lacking is the experimental validation of the computational method development. 

Hopefully, by providing an overview of deep learning methods and detailed explanations of new approaches in the space, we can encourage the more experimentally-minded to utilize methods in this space, or at least to collaborate more confidently. In the other direction, it is important to convey to the methodologically-inclined that while it is great to develop a new method that performs well on a benchmark, experimental validation is the gold standard, so it is beneficial to seek out experimental collaboration when possible. This will lead to more actionable results and push the development of machine learning VS methods even faster in the future.

\section{Acknowledgements}
\label{sec:acknowledgements}

Support for this work was provided by the Office of the Vice Chancellor for Research and Graduate Education at the University of Wisconsin-Madison with funding from the Wisconsin Alumni Research Foundation. In addition, we are thankful for the support provided by the NLM Biomedical Informatics Training Program (grant number 5T15LM007359). Also, we would like to thank Kylie Moynihan for helpful feedback on the manuscript.

\bibliographystyle{vancouver}

\bibliography{cas-refs}

\end{document}